\documentclass[fleqn,usenatbib]{mnras}

\usepackage{newtxtext,newtxmath}

\usepackage[T1]{fontenc}
\usepackage{ae,aecompl}

\usepackage{graphicx}	
\usepackage{amsmath}	
\usepackage{pdflscape}
\usepackage{soul}
\usepackage[normalem]{ulem}

\makeatletter\ifdefined\Hy@StartlinkName
\def\addOneNestingLevelStartLink{%
  \gdef\Hy@StartlinkName##1##2{%
    \sbox0{\Hy@StartlinkNameOrig{##1}{##2}}\usebox0
    \global\let\Hy@StartlinkName\Hy@StartlinkNameOrig%
  }%
}
\def\addOneNestingLevelEndLink{%
  \gdef\pdfendlink{%
    \sbox0{\pdfendlinkOrig}\usebox0%
    \global\let\pdfendlink\pdfendlinkOrig%
  }%
}
\let\Hy@StartlinkNameOrig\Hy@StartlinkName
\let\pdfendlinkOrig\pdfendlink
\else
\let\addOneNestingLevelStartLink\relax
\let\addOneNestingLevelEndLink\relax
\fi

\newcommand{\pkg}[1]{\textsc{#1}}

\newcommand{\Msun}{\, \mathrm{M}_\mathrm{\odot}}
\newcommand{\Msunh}{\, \Msun h^{-1}}

\newcommand{\Mstar}{M_\mathrm{*}}
\newcommand{\Mgas}{M_\mathrm{gas}}
\newcommand{\Mvir}{M_\mathrm{vir}}

\newcommand{\Omegab}{\Omega_\mathrm{b}}
\newcommand{\Omegam}{\Omega_\mathrm{m}}

\newcommand{\Mdotbh}{\dot{M}_\mathrm{BH}}
\newcommand{\Mdotin}{\dot{M}_\mathrm{inflow}}
\newcommand{\Mdotacc}{\dot{M}_\mathrm{acc}}

\newcommand{\MdotEdd}{\dot{M}_\mathrm{Edd}}
\newcommand{\Mdotjet}{\dot{M}_\mathrm{jet}}
\newcommand{\Mdotwind}{\dot{M}_\mathrm{wind}}
\newcommand{\Ledd}{L_\mathrm{Edd}}
\newcommand{\upbound}{R_\mathrm{upper}}
\newcommand{\lowbound}{R_\mathrm{lower}}
\newcommand{\vwind}{v_\mathrm{wind}}
\newcommand{\vwindq}{v_\mathrm{wind,quasar}}
\newcommand{\vwindsd}{v_\mathrm{wind,slim}}

\newcommand{\vjet}{v_\mathrm{jet}}

\newcommand{\epssd}{\epsilon_\mathrm{f,slim}}
\newcommand{\epsq}{\epsilon_\mathrm{f,quasar}}
\newcommand{\epsadaf}{\epsilon_\mathrm{f,ADAF}}
\newcommand{\psiq}{\psi_\mathrm{quasar}(j)}
\newcommand{\psisd}{\psi_\mathrm{slim}(j, \Mdotacc)}
\newcommand{\psiadaf}{\psi_\mathrm{ADAF}}
\newcommand{\psijet}{\psi_\mathrm{jet,res}(j)}
\newcommand{\psijetsub}{\psi_\mathrm{jet}(j)}
\newcommand{\kms}{\mathrm{km}\,\mathrm{s}^{-1}}

\newcommand{\Mbh}{M_\mathrm{BH}}

\newcommand{\cMpch}{\, \mathrm{cMpc}\, h^{-1}}
\newcommand{\invMpccubed}{\mathrm{cMpc}^{-3}}
\newcommand{\Mpccubed}{\mathrm{cMpc}^{3}}
\newcommand{\kpch}{\, \mathrm{kpc}\, h^{-1}}
\newcommand{\facc}{f_\mathrm{acc}}

\newcommand{\mvec}[1]{\boldsymbol{#1}}
\newcommand{\Madaf}{M_\mathrm{ADAF,limit}}
\newcommand{\epscold}{\epsilon_\mathrm{cold}}

\newcommand{\Rfivehundred}{R_\mathrm{500}}
\newcommand{\Mfivehundred}{M_\mathrm{500}}
\newcommand{\cMpc}{\mathrm{cMpc}}
\newcommand{\kpc}{\mathrm{kpc}}
\newcommand{\pc}{\mathrm{pc}}
\newcommand{\etadisc}{\eta_\mathrm{disc}}
\newcommand{\Mbhsub}{M_\mathrm{BH,sub}}
\newcommand{\Mbhdyn}{M_\mathrm{BH,dyn}}

\title[The \textsc{Obsidian} sub-grid model]{The \textsc{Obsidian} model: Three regimes of black hole feedback}

\author[D. Rennehan et al.]{
Douglas Rennehan,$^{1,2}$\thanks{E-mail: douglas.rennehan@gmail.com (DR)}
Arif Babul,$^{2,3}$
Belaid Moa,$^{4}$
and Romeel Dav\'{e}$^{5,6}$
\\
$^{1}$Center for Computational Astrophysics, Flatiron Institute, 162 Fifth Avenue, New York, NY, 10010, USA\\
$^{2}$Department of Physics \& Astronomy, University of Victoria, BC V8X 4M6, Canada\\
$^{3}$Infosys Visiting Chair Professor, Indian Institute of Science, Bangalore 560012, India\\
$^{4}$Department of Electrical Engineering/Alliance Canada/WestDRI/University Systems, University of Victoria, BC V8P 5C2, Canada\\
$^{5}$Institute for Astronomy, University of Edinburgh, Edinburgh EH9 3HJ, United Kingdom\\
$^{6}$Department of Physics and Astronomy, University of the Western Cape, Bellville 7535, South Africa\\
}

\date{Accepted XXX. Received YYY; in original form ZZZ}

\pubyear{2023}

\begin{document}
\label{firstpage}
\pagerange{\pageref{firstpage}--\pageref{lastpage}}
\maketitle

\begin{abstract}
In theoretical models of galaxy evolution, black hole feedback is a necessary ingredient in order to explain the observed exponential decline in number density of massive galaxies.  Most contemporary black hole feedback models in cosmological simulations rely on a constant radiative efficiency (usually $\eta \sim 0.1$) at all black hole accretion rates.  We present the \textsc{Obsidian} sub-grid model, a synthesis model for the spin-dependent radiative efficiencies of three physical accretion rate regimes, i.e. $\eta = \eta(j, \Mdotacc)$, for use in large-volume cosmological simulations.  The three regimes include: an advection dominated accretion flow ($\Mdotacc < 0.03\,\MdotEdd$), a quasar-like mode ($0.03 < \Mdotacc / \MdotEdd < 0.3$), and a slim disc mode ($\Mdotacc > 0.3\,\MdotEdd$).  Additionally, we include a large-scale powerful jet at low accretion rates.  The black hole feedback model we present is a kinetic model that prescribes mass loadings but could be used in thermal models directly using the radiative efficiency.  We implement the \textsc{Obsidian} model into the \pkg{Simba} galaxy evolution model to determine if it is possible to reproduce galaxy populations successfully, and provide a first calibration for further study.  Using a $2\times1024^3$ particle cosmological simulation in a $(150\,\cMpc)^3$ volume, we found that the model is successful in reproducing the galaxy stellar mass function, black hole mass-stellar mass relationship, and stellar mass-halo mass relationship.  Moving forward, this model opens new avenues for exploration of the impact of black hole feedback on galactic environments.
\end{abstract}




\section{Introduction}

The details of the physical processes that halt star formation in galaxies (i.e. the \textit{quenching} process) is an open, long-standing question. The foremost candidate to explain the quenching process in galaxies is active galactic nuclei (AGN) feedback, driven by accretion onto a supermassive black hole (SMBH) \citep{Cattaneo2009}.  There is ample observational evidence that every massive galaxy hosts a central SMBH \citep{Kormendy2013, Bentz2018, Schutte2019, Ding2020}.  Combined with the evidence from large-scale galaxy surveys that show AGN-hosting galaxies have significantly lower star formation rates (SFRs) than their non-AGN counterparts \citep{Ellison2016, Smith2016, Sanchez2018, Lacerda2020, Ellison2021, Lammers2023}, it is clear that theoretical models of galaxy evolution must include accurate models for the impact of AGN feedback.

SMBHs play an outsized role in galaxy evolution through their extremely efficient conversion of accretion energy into powerful feedback in their environments.  That feedback energy must couple to the gaseous environment in order to prevent star formation.  There are two important mechanisms necessary for this to occur.  First, AGN feedback is brilliantly on display in the cores of galaxy groups and clusters, in the form of large-scale ($\sim1 - 10$ $\kpc$) jets \citep{Jetha2008, OSullivan2012, Kuzmicz2017, Brienza2023, Chavan2023}.  These powerful jets ($\sim 50,000$ $\kms$) inflate and heat large, radio-emitting bubbles (i.e. the \textit{radio mode} of feedback) in the intracluster medium \citep{Boehringer1993, Nusser2006, Blanton2011, Babul2013, Prasad2015, Prasad2018, Cielo2018}, which may heat the gas through shocks, sound waves, turbulence, or thermal conduction -- preventing gas cooling onto the central galaxy \citep{Reynolds2005, Conroy2008, Ruszkowski2011, Bourne2021, Su2021, Bourne2023, Weinberger2023}.  Second, other galaxies could be more impacted by quasar radiation on the galaxy-scale (i.e. the \textit{quasar mode} of feedback) rather than jets directly \citep{Babul1991, Silk1998, King2015}.  The quasar winds are thought to be launched by radiation pressure on gas close to the SMBH from the accretion process itself \citep{Murray2005}.  They should escape the central region preferentially along the path of least resistance, perpendicular with some opening angle to the galactic disc with velocities $\sim1000$ $\kms$ on $\sim \kpc$ scales \citep{Ciotti2009, Faucher-Giguere2012, Fiore2017}.  Observations on smaller scales show the existence of ultra-fast outflows with velocities up to $\sim50,000$ $\kms$ in the cores of $\approx20\%$ to $40\%$ of local AGN \citep{Tombesi2010, Tombesi2011, Gofford2013}.  Even a small amount of the AGN feedback coupling to the surrounding interstellar or circumgalactic medium should be sufficient to halt star formation in most galaxies.

Save for large-scale collimated jets in galaxy clusters, the accretion and feedback processes within AGN occur on scales close to the SMBH ($\sim 0.1-100$ $\pc$).  Contemporary cosmological simulations with SMBHs have made significant gains in resolution and volume (see e.g. \citealt{AnglesAlcazar2021, Koudmani2023, Hopkins2023b}), however even the most ambitious large-scale simulations achieve effective resolutions on the $\kpc$-scale.  Therefore, capturing the complete physics of the AGN while simultaneously simulating thousands of galaxies would require an infeasible increase of a factor of up to $\sim10^4$ in computation time. 
 For that reason, AGN related physics must be treated with sub-grid scale models -- models that approximate the small-scale physics and link to the resolved scale (for excellent reviews see \citealt{Somerville2015, Naab2016, Crain2023}).   

The first sub-grid models for AGN feedback in galaxy-scale simulations were highly simplified, but were able to show that merging disc galaxies power an AGN that is able to quench star formation \citep{Springel2005b, DiMatteo2005, DiMatteo2007a}.  Most contemporary models use a similar approach, where a small, tunable fraction $\epsilon_\mathrm{f} \sim 0.05$ of the bolometric luminosity of the AGN couples to medium, providing all of the necessary energy.  Typically, the luminosity is paramterised as $L = \eta \Mdotacc c^2$, with $\eta$ being the radiative efficiency and $\Mdotacc$ the accretion rate onto the black hole \citep{Soltan1982}.  Since these models were originally motivated by quasars in post-merger galaxies, the radiative efficiency is usually taken as a constant $\eta \sim 0.1$.  Indeed, quasar mode feedback is effective at clearing out the gas in the cores of simulated galaxies, which regulates the SFR in the galaxy, as well as preventing the SMBH from growing too rapidly \citep{Silk1998, Sazonov2005, Cattaneo2009}, explaining the relationship between SMBH mass and central velocity dispersion \citep{Kormendy2013}. While the quasar mode is effective in lower mass halos, the radio mode is necessary in massive galaxies \citep{Nemmen2007} at the galaxy group and cluster scale \citep{McCarthy2008, Nusser2006, Cielo2018, Oppenheimer2021, Jung2022}.

There are a plethora of inventive ways to actually implement these energetics into cosmological simulations.  However, AGN feedback models depend on the resolution, hydrodynamical method, and pure philosophical choice.  There have been several large-volume simulations, with volumes greater than $(100\, \cMpc)^3$, simulated over the last decade with different AGN feedback models.  The original \pkg{Illustris} simulation modelled AGN feedback by thermally dumping energy directly in the gas surrounding SMBHs at high accretion rates, while producing radio bubbles directly in the halo gas on large scales \citep{Sijacki2007, Vogelsberger2014, Sijacki2015a}.  In subsequent work, \pkg{IllustrisTNG} (The Next Generation) modelled the low accretion rate regime with a kinetic wind rather than radio bubbles \citep{Weinberger2016, Pillepich2018}.  The \pkg{EAGLE} simulations used a similar style of feedback to \pkg{Illustris}, where the simulated SMBHs dumped thermal energy locally in the surrounding gas, although with a reservoir that stored energy until it could overcome the artificially high cooling rates in the simulated interstellar medium \citep{Booth2009, DallaVecchia2012, Rosas-Guevara2015a, Schaye2014}.  \pkg{Horizon-AGN} introduced a similar split in accretion rate ratios, where rapidly accreting SMBHs dumped thermal energy directly to the surrounding gas \citep{Dubois2014, Kaviraj2017}.  At low accretion rates, the SMBHs ejected a bipolar jet with a $10\%$ efficiency \citep{Dubois2012}.  Contrary to the popular thermal dump, the \pkg{Simba} simulations convert the simulated SMBH luminosities to kinetic power, in order to power kinetic winds \citep{Dave2019}.  At low accretion rate ratios, they use a powerful kinetic jet that ramps up in power as the SMBH accretion rate drops.  Recently, the \pkg{FLAMINGO} simulations reintroduced a model similar to the original \pkg{EAGLE} simulations, but added a kinetic jet with very low efficiency of $1.5\%$ at low accretion rates in some of their non-fiducial tests \citep{Husko2022, Schaye2023}.

There have been several improvements to AGN feedback models in zoom-in simulations, which focus on high-resolution simulations of individual halos.  For example, \pkg{Simba} was recalibrated for \pkg{The 300} simulations \citep{Cui2022}, \pkg{EAGLE} recalibrated for the \pkg{C-EAGLE} \citep{Barnes2017} and \pkg{BAHAMAS} \citep{McCarthy2017} suites of galaxy clusters, \pkg{Horizon-AGN} updated to the \pkg{NewHorizon} simulation \citep{Dubois2021}, and \pkg{Illustris} recalibrated and updated for the \pkg{FABLE} simulations \citep{Henden2018}.  Similarly, the \pkg{Romulus} suite \citep{Tremmel2017a} and the \pkg{FIRE} suite \citep{Hopkins2014, Hopkins2018, Hopkins2023} are focused, high-resolution simulations that include AGN feedback models.  In both cases, they use quasar-like feedback although the \pkg{FIRE} simulations have dual mode thermal/kinetic injection of energy into the gas surrounding their simulated SMBHs \citep{Wellons2023}, whereas \pkg{Romulus} uses a single thermal dump.

Large-volume cosmological simulations that include AGN feedback have been able to reproduce the downturn in the galaxy stellar luminosity function by dropping the star formation efficiency in massive galaxies, thereby quenching them.   While these models are impressive feats, all of the \textit{large-volume} implementations usually model a quasar mode constant radiative efficiency $\eta \sim 0.1$, combined with some radio-mode feedback model at some tunable, low accretion rate.  However, the power available to feedback in the environment depends on both the spin of the SMBH and the accretion rate in a more complicated manner than a single threshold, and it is important to revisit this as it is a key parameter in determining the overall physics of the system.   

SMBHs are characterized by three parameters: mass $\Mbh$, angular momentum $J$, and charge -- although they typically have zero charge \citep{Blandford1977}.  The angular momentum is normally recast into the dimensionless spin, $j$, of the SMBH via $j = Jc/G\Mbh^2$, where $c$ is the speed of light and $G$ is Newton's gravitational constant.  The dimensionless spin is restricted to the range $-1 < j < 1$, and SMBHs may change their angular momentum through mergers with other SMBHs or through accretion \citep{Bardeen1975, Scheuer1996, Gammie2004, Berti2008, Reynolds2012, Babul2013, Ricarte2023}.  The geometry of accretion flows is complex, but the standard models usually involve magnetised discs and hot accretion flows at low accretion rates, or combinations thereof.  The radiative and jet efficiencies of the accretion process depends on the structure of the discs and, in particular, the innermost, spin-dependent, boundary condition near the SMBH \citep{Bardeen1970, Thorne1974, Gammie1999, Narayan2003}.  Although the spin distribution of SMBHs is uncertain, the radiative and jet efficiencies usually increase with the spin of the SMBH (see e.g. \citealt{Nemmen2007, Benson2009, Reynolds2012, Tchekhovskoy2012}).

There are three accretion disc models, separated on the basis of the accretion rate (or luminosity): (i) the advection dominated accretion flow (ADAF), (ii) thin disc, and (iii) slim disc models.

The ADAF model describes a system of a geometrically thick, optically thin hot gas that may be combined with a truncated thin disc at some outer radius (a \textit{hot} accretion flow) \citep{Yuan2014}.  At very low accretion rates ($\Mdotacc \lesssim 0.03 \MdotEdd$), the flow is hot and unable to cool rapidly causing energy to be advected into the black hole \citep{Esin1997}.   This difference in long cooling times versus short in-fall times leads to the flow to be radiatively inefficient, with $\eta$ scaling as $\eta \sim \Mdotacc / \MdotEdd$.  The model has been successful at reproducing the spectral properties of low-luminosity active galactic nuclei (AGN) including the supermassive black hole in the center of our galaxy.  In this regime, it is possible to have a powerful kinetic jet combined with a wind generated from the accretion flow \citep{Benson2009}.

Accretion flows around black holes which accrete more rapidly ($0.03 \lesssim \Mdotacc / \MdotEdd \lesssim 0.3$) are better described by a thin disc model, often modelled as a Shakura-Sunyaev viscous thin disc \citep{Shakura1973}.  In this case, the disc is both geometrically thin and optically thick, as the disc is in local thermal equilibrium and can radiate its viscous heat efficiently \citep{Laor1989}.  Here the radiative efficiency is the highest, with $\eta \sim 0.1$ for a non-rotating SMBH \citep{Shakura1973}, although it can be much higher for a rotating SMBH \citep{Bardeen1970}.  The model has successfully been used to explain the spectra of quasars which have a notable blackbody feature.

At very high accretion rates ($\Mdotacc \gtrsim 0.3 \MdotEdd$) the flow becomes optically thick and geometrically thin, but much more optically thick than the thin disc model \citep{Abramowicz1988, Sadowski2014}.  The dissipated energy cannot be radiated rapidly enough and can be advected with the flow leading to a drop in the radiative efficiency ($\eta \sim [\Mdotacc / \MdotEdd]^{-1}$) (\citealt{Sadowski2009, Madau2014a}; see e.g. \citealt{Lupi2016} for an implementation in a cosmological context).  The dissipated energy causes a slight thickening of the disc and, therefore, this model has been named the slim disc model \citep{Czerny2019}.

In each of the accretion disc models, the spin impacts the radiative efficiencies directly.  The subsequent feedback into the environment of SMBHs is not only dependent on the spin value, but the direction of the spin \citep{Babul2013, Cenci2020, Sala2020}.  In large-volume cosmological simulations, the resolution is usually too poor to follow the spin-evolution of SMBHs (however, see e.g. \citealt{Dubois2021, Schaye2023}).  Idealised simulations or cosmological zoom-in simulations provide the ideal test beds for tracking the spin of the SMBH and its impact on the radiative efficiencies and, hence, the SMBH feedback strength (see e.g., \citealt{Fiacconi2018, Qiu2019, Koudmani2023, Talbot2022, Talbot2024}).  There has been a particular focus on spin and accretion rate dependent radiative efficiencies in the ADAF mode, where the spin dependence can cause spin precession that drives isotropic turbulence and prevent cooling in the cores of galaxy clusters \citep{Beckmann2019, Beckmann2022}.

The goal of this work is to incorporate the radiative efficiency scalings (with respect to the accretion rate) from the high-resolution simulations into coarser, large-volume cosmological simulations in a consistent manner.  Specifically, we aim to synthesise the disc wind and jet model from \cite{Benson2009}, the slim disc radiative efficiencies from \cite{Sadowski2014}, and the standard $\eta \sim 0.1$ quasar-like mode of feedback into a cosmological-scale simulation.  Our goal in this work is not to follow the spin evolution of the SMBHs. This paper is organised as follows. First, in Section~\ref{sec:new_model}, we describe the new radiative efficiency model in full mathematical detail along with a kinetic injection mechanism. In Section~\ref{sec:methods}, we describe our motivation for incorporating our model into the \pkg{Simba} model, and give details of the implementation.  In Section~\ref{sec:calibration}, we discuss the calibration of the newly incorporated model.  In Section~\ref{sec:galaxies}, we describe the simulated population of galaxies and broad results.  Finally, we summarize our findings in Section~\ref{sec:conclusions}.

\section{Physically-motivated black hole feedback}
\label{sec:new_model}

We present a finite state machine for treating supermassive blackhole (SMBH) feedback in cosmological simulations.  There are three physical regimes which we must consider:

\begin{enumerate}
    \item $\Mdotacc / \MdotEdd > \upbound$,
    \item $\lowbound < \Mdotacc / \MdotEdd \leqslant \upbound$, and
    \item $\Mdotacc / \MdotEdd \leqslant \lowbound$,
\end{enumerate}

\noindent where $\upbound$ is the boundary between the slim disc mode and the traditional quasar mode and $\lowbound$ is the boundary between the advection-dominated accretion flow (ADAF) mode and quasar mode.  In our notation, the true growth rate of the SMBH is the accretion rate $\Mdotacc$ accounting for radiative losses,

\begin{equation}
    \label{eq:true_accretion_rate}
    \Mdotbh = (1 - \eta(j,\Mdotacc))\Mdotacc.
\end{equation}

Note that $\MdotEdd$ is the Eddington accretion rate,

\begin{equation}
    \label{eq:eddington_accretion_rate}
    \MdotEdd \equiv \frac{4\pi G m_\mathrm{p}}{\sigma_\mathrm{T}\eta_\mathrm{fixed} c} M_\mathrm{BH}.
\end{equation}

\noindent Here $G$ is Newton's gravitational constant, $m_\mathrm{p}$ is the proton mass, $\sigma_\mathrm{T}$ is the Thomson cross section, $c$ is the speed of light, $\eta_\mathrm{fixed} = 0.1$ is a normalisation radiative efficiency, and $M_\mathrm{BH}$ is the mass of the black hole.

We treat each regime as a unique state in which the SMBH exists until the conditions are met to transfer to a new state.  The transfer condition is dependent on the current state of the SMBH through the \textit{true accretion rate} onto the SMBH ($\Mdotacc$) not the \textit{large-scale mass inflow rate} ($\Mdotin$).  Note that in our model, the transfer between states is instantaneous.

While the boundaries $\upbound$ and $\lowbound$ are only approximate, we fix them to values $\upbound = 0.3$ and $\lowbound = 0.03$, in line with theoretical estimates \citep{Laor1989, Maccarone2003, Greene2006, McClintock2006, Sadowski2009, Madau2014a}.

\begin{figure}
    \centering
    \includegraphics[scale=0.65]{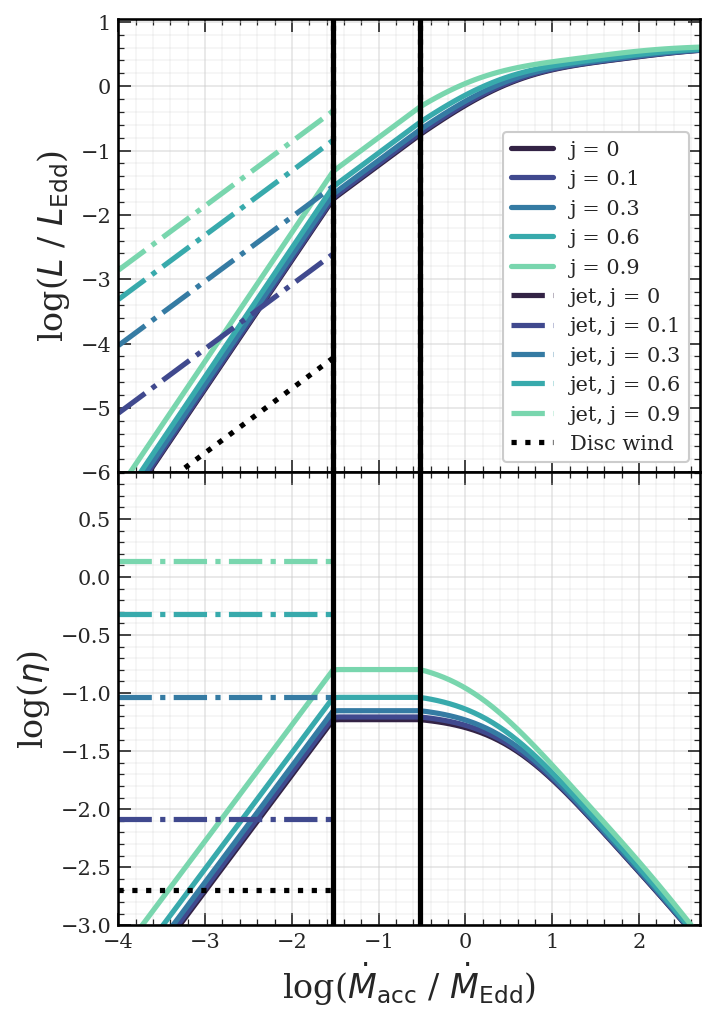}
    \caption{(top) The black hole bolometric luminosity as a function of the small-scale accretion rate.  (bottom) The radiative efficiency as a function of the small-scale accretion rate.  In both panels, the line colours show the black hole spin, $j$, increasing from dark to light, respectively.  The solid lines show the small scale feedback luminosity and the dash-dotted lines show the jet power. The dashed vertical lines show the boundaries between the three accretion regimes.  As black hole spin increases, the black hole draws more power from the accretion flow and the jet becomes significantly more powerful.  The typical feedback model uses a radiative efficiency $\eta \sim 0.1$ which corresponds to the middle accretion rate regime.}
    \label{fig:agn_power_eta}
\end{figure}

In Fig.~\ref{fig:agn_power_eta}, we show the bolometric luminosity (top) and spin-dependent radiative efficiencies (bottom) that we use throughout this work.  In both panels, the solid lines show the values for the non-jet component, whereas the dash-dotted lines show the jet component.  The dotted vertical lines show the thresholds between the three regimes of feedback, depending on the accretion rate ratio $\Mdotacc / \MdotEdd$.  We produced the top panel directly from the bottom via the equation $L = \eta(j, \Mdotacc) \Mdotacc c^2$.  The $\eta$ scalings for the low accretion rate and high accretion rate regime were taken from \cite{Benson2009} and \cite{Madau2014a}, respectively, and the normalisation was found by ensuring continuity in the radiative efficiency in all regimes (starting with the high accretion rate regime).  We discuss the radiative efficiency functions in the following sub-sections.

\subsection{Slim disc mode}
\label{sec:new_model_slim_disc}

For the high accretion rate regime we follow \cite{Lupi2016} and use the radiative efficiency from \cite{Sadowski2014}, and \cite{Madau2014a}.

\begin{equation}
    \eta_\mathrm{high}(j, r) = \frac{r}{16} A(j) \bigg[ \frac{0.985}{r + B(j)} + \frac{0.015}{r + C(j)}\bigg]
    \label{eq:high_edd_efficiency},
\end{equation}

\noindent where $j$ is the black hole spin parameter and $r \equiv \MdotEdd^{*} / \Mdotacc$.  Note $\MdotEdd^* \neq \MdotEdd$; it is defined as $\MdotEdd^* \equiv 16\, \Ledd / c^2$ or $\MdotEdd^* = 1.6\, \MdotEdd$.  Additionally,

\begin{equation}
    A(j) \equiv (0.9663-0.9292j)^{-0.5639},
    \label{eq:aj_definition}
\end{equation}
\begin{equation}
    B(j) \equiv (4.627-4.445j)^{-0.5524},
    \label{eq:bj_definition}
\end{equation}

\noindent and

\begin{equation}
    C(j) \equiv (827.3-718.1j)^{-0.7060}.
    \label{eq:cj_definition}
\end{equation}

The accretion rate onto the supermassive black hole $\Mdotacc$ is the difference between the large scale mass flow rate and the outflowing wind

\begin{equation}
    \label{eq:app_slim_disc_1}
    \Mdotacc = \Mdotin - \Mdotwind = \Mdotin - \psisd\Mdotacc,
\end{equation}

\noindent where we assume a mass loading of $\psisd$ between $\Mdotacc$ and the wind mass outflow rate, $\Mdotwind$ \citep{Choi2012a, Choi2015a, Choi2017}.  Following the approach in \cite{Dave2019}, we assume the SMBH powers a wind via a momentum loading \citep{Murray2005}

\begin{equation}
    \label{eq:app_slim_disc_2}
    \Mdotwind\vwind = \frac{20\,L_\mathrm{BH}}{c} = 20\, \eta(j,\Mdotacc)\Mdotacc c,
\end{equation}

\noindent where $\Mdotwind$ is the mass outflow rate, $\vwind$ is the wind velocity, and $\eta(j,\Mdotacc)$ is the radiative efficiency.  We assume $\sim 20$ following \cite{Faucher-Giguere2012} and \cite{Zubovas2012}. Therefore,

\begin{equation}
    \label{eq:app_slim_disc_3}
    \psisd = \frac{\Mdotwind}{\Mdotacc} = \epssd \frac{20\,\eta(j,\Mdotacc) c}{\vwind} \equiv \phi\eta(j,\Mdotacc),
\end{equation}

\noindent where we have further defined $\phi \equiv 20c\epssd/\vwind$ in order to show the explicit dependence on $\eta$.  We introduce a parameter $\epssd$ to control the momentum flux of the wind since the resolution of our simulations cannot accurately  capture the hydrodynamical interaction of the wind with the surrounding gaseous medium.  We treat both $\epssd$ and $\vwind$ as free parameters and we discuss their calibration in Section~\ref{sec:calibration}.

We further define the following accretion rate ratios for $\Mdotacc$ and $\Mdotin$:

\begin{equation}
    \label{eq:app_slim_disc_4}
    \mathcal{R} \equiv \frac{\Mdotacc}{\MdotEdd},\\ \mathcal{R}' \equiv \frac{\Mdotin}{\MdotEdd}.
\end{equation}

\begin{figure}
    \centering
    \includegraphics[scale=0.65]{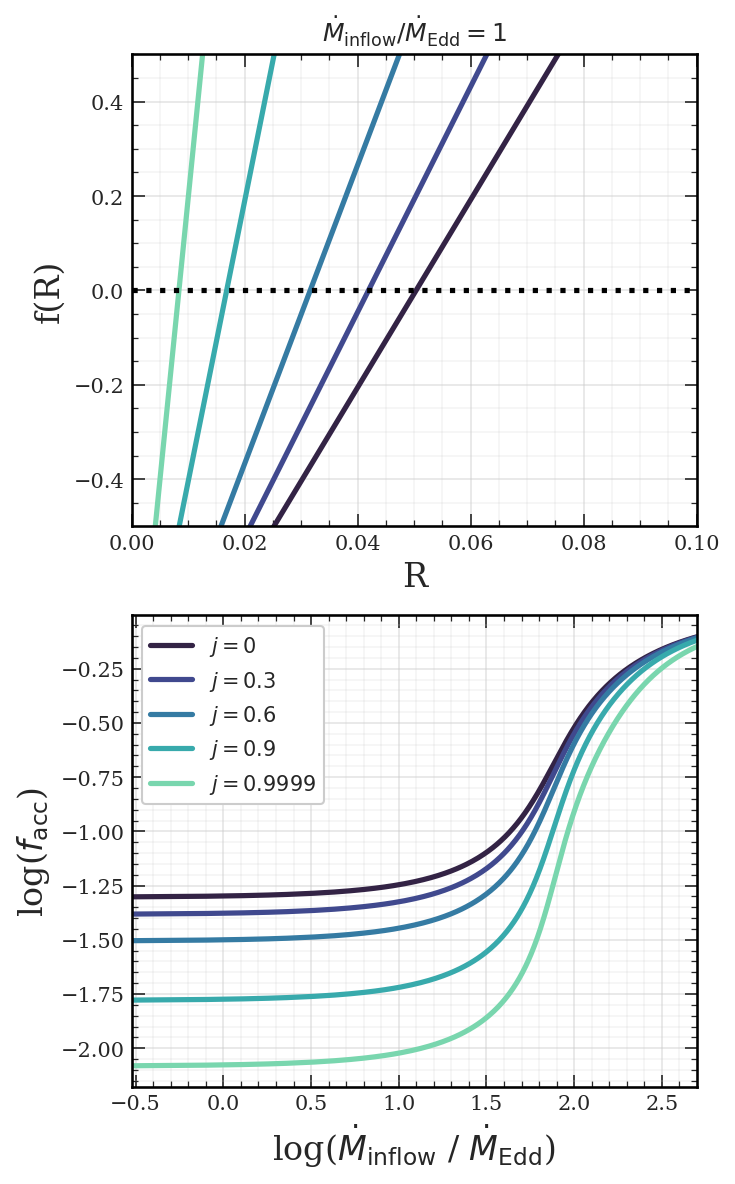}
    \caption{(top) Examples of the cubic function $f(\mathcal{R})$ for the slim disc mode in equation~\ref{eq:app_slim_disc_7}, with fixed $\Mdotin/\MdotEdd = 1$.  The horizontal dashed line shows the zero point of the function, where we solve the cubic equation for $\mathcal{R} \equiv \Mdotacc/\MdotEdd$, giving the true accretion rate onto the black hole.  (bottom) The accretion fraction, i.e. $\facc\equiv\Mdotacc/\Mdotin$, for the slim disc mode as a function of the large-scale inflow rate $\Mdotin/\MdotEdd$.  In both panels, the coloured lines show different choices for the black hole spin for low spins (dark) to high spins (light).  Higher values of $j$ lead to lower accretion fractions, implying that the black hole has more mass in outflows while simultaneously being more likely to switch from the slim disc mode to the quasar mode (i.e. once $\mathcal{R} = \Mdotacc/\MdotEdd < 0.3$).}
    \label{fig:super_eddington_accretion}
\end{figure}

Using equation~\ref{eq:high_edd_efficiency} for $\eta$, we rewrite equation~\ref{eq:app_slim_disc_1} as,

\begin{equation}
    \label{eq:app_slim_disc_5}
    \mathcal{R} + \frac{\phi}{16}A(j)\bigg[\frac{0.985\mathcal{R}}{1 + (5/8) B(j)\mathcal{R}}+\frac{0.015\mathcal{R}}{1 + (5/8)C(j)\mathcal{R}}\bigg] = \mathcal{R}'.
\end{equation}

\noindent We further simplify this equation by multiplying the entire equation by $[1 + (5/8)B(j)\mathcal{R}][1 + (5/8)C(j)\mathcal{R}] = 1 + (5/8)(B(j) + C(j))\mathcal{R} + (5/8)^2B(j)C(j)\mathcal{R}^2$ and reducing to obtain

\begin{equation}
    \label{eq:app_slim_disc_6}
    \begin{split}
    \mathcal{R}+\frac{5}{8}(B(j)+C(j))\mathcal{R}^2 + \bigg(\frac{5}{8}\bigg)^2B(j)C(j)\mathcal{R}^3 + \\
    \frac{\phi}{16}A(j)\bigg(\mathcal{R} + \frac{5}{8}[0.015B(j) + 0.985C(j)]\mathcal{R}^2\bigg)\\
    =\mathcal{R}' + \frac{5}{8}(B(j)+C(j))\mathcal{R}\mathcal{R}' + \bigg(\frac{5}{8}\bigg)^2B(j)C(j)\mathcal{R}^2\mathcal{R}'.
    \end{split}
\end{equation}

\noindent This is a cubic equation in $\mathcal{R}$ (i.e. $\Mdotacc / \MdotEdd$) since we know $B(j)$, $C(j)$, $\MdotEdd$, and $\mathcal{R}'$ (i.e. $\Mdotin / \MdotEdd$) at simulation time.  Reducing the equation,

\begin{equation}
    \label{eq:app_slim_disc_7}
        f(\mathcal{R}) = \alpha_3 \mathcal{R}^3 + \alpha_2 \mathcal{R}^2 + \alpha_1 \mathcal{R} + \alpha_0 = 0,
\end{equation}

\noindent where we have defined

\begin{equation}
    \label{eq:app_slim_disc_8}
    \begin{split}
        \alpha_3 \equiv \bigg(\frac{5}{8}\bigg)^2B(j)C(j),\\
        \alpha_2 \equiv \frac{5}{8}\bigg[B(j) + C(j) + \frac{\phi}{16}A(j)(0.015 B(j) \\
        + 0.985  C(j)) - \frac{5}{8} B(j) C(j) \mathcal{R}'\bigg],\\
        \alpha_1 \equiv 1 + \frac{\phi}{16}A(j) - \frac{5}{8}(B(j) + C(j))\mathcal{R}', \mathrm{and}\\
        \alpha_0 \equiv -\mathcal{R}'.
    \end{split}
\end{equation}

In the top panel of Fig.~\ref{fig:super_eddington_accretion}, we show the function $f(\mathcal{R})$ from equation~\ref{eq:app_slim_disc_7} for values of $j$, $\phi$, and $\mathcal{R}'$.  The lines show spin values $j \in \{0, 0.3, 0.6, 0.9, 0.9999\}$ from darkest to lightest, respectively, and we fix $\Mdotin/\MdotEdd = 1$ as well as $\phi$ using values $v_\mathrm{wind} = 1000$ $\kms$ and $\epssd = 1$.  The dashed line shows $f(\mathcal{R}) = 0$, and the solution for the accretion rate $\mathcal{R}$ lies at the intersection of the solid lines and the dashed line.  For this illustrative purpose (a black hole with spin $j$ accreting at the Eddington limit), the predicted true accretion rate $\Mdotacc$ is much less than the large-scale inflow rate $\Mdotin$, indicating that a large fraction of $\Mdotin$ is lost in the form of an outflowing wind.  In this case, the black hole would not remain in the slim disc mode, and would transition to a new state. 

The general behaviour in the slim disc regime is a drop off in radiative efficiency as $\eta \sim (\Mdotacc / \MdotEdd)^{-1}$ (recall Fig.~\ref{fig:agn_power_eta}).  Our model includes a radiatively driven wind, with only a fraction of $\Mdotin$ entering the SMBH at a rate $\Mdotacc$,

\begin{equation}
    \label{eq:app_slim_disc_9}
    \Mdotacc = \bigg(\frac{1}{1+\psisd}\bigg)\Mdotin.
\end{equation}

\noindent Accounting for the radiative losses, the true mass growth rate of the BH is given by

\begin{equation}
    \label{eq:slim_disc_mdotbh}
    \Mdotbh = \bigg(\frac{1 - \eta(j, \Mdotacc)}{1 + \psisd}\bigg)\Mdotin.
\end{equation}

\noindent The bottom panel of Fig.~\ref{fig:super_eddington_accretion} shows the behaviour of $\facc\equiv\Mdotacc/\Mdotin$ as a function of the large-scale accretion rate $\Mdotin / \MdotEdd$.  The lines are coloured by the spin $j$, just as the top panel.  There is a factor of $\sim 100$ difference in the accretion fractions depending on the large-scale inflow rate, indicating that, at the highest inflow rates ($\Mdotin / \MdotEdd \gtrsim 10^2$), the black holes should be able to grow very rapidly due to a lack of a radiatively-driven outflowing wind.  Note that this behaviour depends on $v_\mathrm{wind}$ and $\epssd$ and, therefore, these parameters directly control the growth speed of the black holes in this regime.

\subsection{Quasar mode}
\label{sec:new_model_quasar}

We treat the energetics in the quasar mode identically as in the slim disc mode, except with a radiative efficiency that only depends on the spin of the black hole, $\eta = \eta(j)$.  To reiterate, the accretion rate onto the supermassive black hole is the difference between the large-scale inflow rate and the mass outflow rate,

\begin{equation}
    \label{eq:quasar_mass_balance}
    \Mdotacc = \Mdotin - \Mdotwind = \Mdotin - \psiq\Mdotacc.
\end{equation}

\noindent Therefore, we have

\begin{equation}
    \label{eq:quasar_mdotacc}
    \Mdotacc = \frac{1}{1+\psiq} \Mdotin.
\end{equation}

Now, we assume that the radiative wind carries a momentum $\sim 20L_\mathrm{BH}/c$

\begin{equation}
    \label{eq:quasar_wind_balance}
    \Mdotwind v_\mathrm{wind} = \epsq \bigg(\dfrac{20L_\mathrm{BH}}{c}\bigg) = \epsq[20\eta(j)\Mdotacc c],
\end{equation}

\noindent where $\eta(j)$ is the radiative efficiency and, like $\epssd$, $\epsq$ is a free parameter that describes the fraction of $20 L / c$ that should be used in the simulation to account for resolution effects.  

\noindent With these results, the energy, momentum, and mass fluxes in the wind are given by 

\begin{equation}
    \label{eq:quasar_energy_flux}
    \dot{E}_\mathrm{wind} = \frac{200(\epsq\eta(j))^2}{\psiq(1+\psiq)} \Mdotin c^2,
\end{equation}

\begin{equation}
    \label{eq:quasar_momentum_flux}
    \dot{P}_\mathrm{wind} = \frac{20\epsq\eta(j)}{1+\psiq} \Mdotin c,
\end{equation}

\noindent and

\begin{equation}
    \label{eq:quasar_mass_flux}
    \Mdotwind = \frac{\psiq}{1+\psiq} \Mdotin,
\end{equation}

\noindent respectively.  Accounting for radiative losses, the true mass growth rate of the black hole is

\begin{equation}
    \label{eq:quasar_mass_growth}
    \Mdotbh = [1 - \eta(j)]\Mdotacc = \frac{1 - \eta(j)}{1+\psiq}\Mdotin.
\end{equation}

\subsection{Advection dominated accretion flow mode}
\label{sec:new_model_adaf}

At the lowest accretion rates, we model both an isotropic feedback component and a jet emanating from the region surrounding the supermassive black hole (SMBH).  For the isotropic component, we assume that it is powered by outflows engendered by magnetic fields in the rotating flow near the SMBH, i.e. a \textit{disc} wind with efficiency $\etadisc$.  For the disc wind power, \cite{Benson2009} suggest a constant value of $\etadisc \sim 0.002$.

We model the jet component via the \cite{Blandford1977} mechanism, where twisted magnetic field lines can extract energy from a rotating SMBH.  The increase in magnetic pressure provides the mechanism to launch the jet, and we follow \cite{Talbot2020} by using their Blandford-Znajek efficiency,

\begin{equation}
    \label{eq:adaf_jet_efficiency}
    \eta_\mathrm{jet}(j) = \frac{\kappa}{4\pi} \phi^2_\mathrm{BH}(j) f(j),
\end{equation}

\noindent where $\kappa = 1/(6\pi)$, $\phi_\mathrm{BH}$ is the dimensionless magnetic flux, and $f(j)$ is some function of the SMBH spin.  For $\phi_\mathrm{BH}$, we follow \cite{Husko2022} and use the results for the spin-dependent magnetic flux from \cite{Narayanan2021},

\begin{equation}
    \label{eq:adaf_jet_phi}
    \phi_\mathrm{BH}(j) = -20.2j^3 - 14.9j^2 + 34j + 52.6.
\end{equation}

\noindent For $f(j)$, we use the result in \cite{Talbot2020} from \cite{Tchekhovskoy2012},

\begin{equation}
    \label{eq:adaf_jet_fj}
    f(j) = \mathcal{J}^2(j) + 1.38\mathcal{J}^4(j) - 9.2\mathcal{J}^6(j),
\end{equation}

\noindent where

\begin{equation}
    \label{eq:adaf_jet_fj_J}
    \mathcal{J} \equiv \frac{j}{2\big(1 + \sqrt{1-j^2}\big)}.
\end{equation}

To model these processes, we first consider mass conservation.  Not all of $\Mdotin$ reaches the SMBH, as some of that material is outflowing as a jet or converted into energy.  That is represented by the mass balance

\begin{equation}
    \label{eq:adaf_first_mass_balance}
    \Mdotacc = \Mdotin - \Mdotjet - \Mdotwind.
\end{equation}

\noindent where $\Mdotjet$ is the mass outflow rate of the jet and $\Mdotwind$ is the mass outflow rate of the isotropic component.  In both cases we assume there is a mass loading rate such that

\begin{equation}
    \label{eq:adaf_second_mass_balance}
    \Mdotacc = \Mdotin - \psijetsub\Mdotacc - \psiadaf\Mdotacc
\end{equation}

\noindent and, therefore,

\begin{equation}
    \label{eq:adaf_third_mass_balance}
    \Mdotacc = \frac{1}{1 + \psijetsub + \psiadaf} \Mdotin.
\end{equation}

The true growth rate of the SMBH must account for radiative losses, as well as the rest-mass energy extracted to power the jet

\begin{equation}
    \label{eq:adaf_mass_balance}
    \Mdotbh = \Mdotacc - \frac{L_\mathrm{rad}}{c^2} - \frac{\dot{E}_\mathrm{jet}}{c^2},
\end{equation}

\noindent where $L_\mathrm{rad} = \eta(j, \Mdotacc)\Mdotacc c^2$, $\dot{E}_\mathrm{jet} = \eta_\mathrm{jet}(j)\Mdotacc c^2$, and $c$ is the speed of light.   Here, $\eta(j, \Mdotacc) \propto \Mdotacc / \MdotEdd$ and we obtain the jet efficiency $\eta_\mathrm{jet}$ from equation 23 in \cite{Talbot2020}.  We substitute these relations into equation~\ref{eq:adaf_mass_balance} to obtain the true accretion rate onto the SMBH in terms of $\Mdotacc$,

\begin{equation}
    \label{eq:adaf_final_mdotacc}
    \Mdotbh = [1 - \eta(j, \Mdotacc) - \eta_\mathrm{jet}(j)] \Mdotacc.
\end{equation}

\noindent Equation~\ref{eq:adaf_third_mass_balance} allows us to link $\Mdotacc$ to the large-scale inflow rate, $\Mdotin$.  First, to obtain our final equation, we must ensure continuity in $\eta$ across all states by scaling the radiative efficiency to the value at $\lowbound$,

\begin{equation}
    \label{eq:adaf_continuity}
    \eta(j, \Mdotacc) = \eta(j, \lowbound)\bigg(\frac{\Mdotacc}{\MdotEdd}\bigg).
\end{equation}

\noindent This is valid for $\Mdotacc/\MdotEdd < \lowbound$.  Finally, combining equations~\ref{eq:adaf_third_mass_balance}, ~\ref{eq:adaf_final_mdotacc}, \& \ref{eq:adaf_continuity} gives the final true SMBH growth rate in terms of $\Mdotin$,

\begin{equation}
    \label{eq:adaf_final_mdotin}
    \Mdotbh = \bigg[\frac{1-\frac{\eta(j, \lowbound)\Mdotin}{(1+\psijetsub+\psiadaf)\MdotEdd}-\eta_\mathrm{jet}(j)}{1+\psijetsub+\psiadaf}\bigg]\Mdotin.
\end{equation}

\noindent All that remains is to determine both $\psijetsub$ and $\psiadaf$.  However, there are a few important considerations before we may determine the mass loadings for the jet and isotropic wind.  

For the jet, one could assume that it conserves either the launch momentum or the launch energy.  On small scales, we expect the jet to approximately conserve energy as it is injected very close to the SMBH.  However, the scales of simulations in this work are much larger than the inner accretion flow.  By the time the jet reaches our resolution in the intracluster medium (approx. $10$ $\kpc$), we would expect that material has been swept up in a larger flow, loaded with momentum.  Indeed, observational techniques to determine the velocities of jets on large-scales show that the advance velocities are $\sim5000-10,000$ $\kms$ \citep{Jamrozy2005, Machalski2007} which is much slower than a typical, expected launch velocity of $\sim50,000$ $\kms$  (see e.g.  \citealt{Tombesi2011, Tombesi2014, Morganti2017}).  We experimented with energy conservation and found that only a narrow range of $\vjet$ was possible and, therefore, we choose to model both a small-scale energy loading $\psijetsub$ combined with a resolved momentum loading $\psijet$.

It is important to emphasise that $\psijetsub$ enters directly into the mass balance through equations~\ref{eq:adaf_third_mass_balance}\,\&~\ref{eq:adaf_final_mdotin}.  Consider the case when $\psiadaf \to 0$, then the accretion fraction of $\Mdotin$ to $\Mdotbh$ is,

\begin{equation}
    \label{eq:adaf_accretion_fraction}
    \facc = \frac{1 - \eta_\mathrm{jet}(j)}{1 + \psijetsub}.
\end{equation}

\noindent In the case that $\eta_\mathrm{jet} > 1$, the jet should be extracting rotational energy from the SMBH thereby reducing the overall mass --- which is the irreducible mass combined with the rotational energy (i.e. $E/c^2$).  Since we do not model the spin-up or spin-down of the SMBH, we also do not subtract $\eta_\mathrm{jet}$ from the accretion fraction to avoid negative SMBH masses in the simulation.  Additionally, $\psijetsub$ is degenerate with the accretion rate normalisation.  For that reason, we fix the small-scale mass loading factor to $\psijetsub = 2000$ at $\eta_\mathrm{jet}(j) = 1$ such that the SMBHs remain on the BH mass-stellar mass relationship.

To compute the resolved mass loading, $\psijet$, we use a momentum constraint assuming that the jet is electromagnetic in origin,

\begin{equation}
    \label{eq:adaf_energy_balance}
    \dot{P}_\mathrm{jet} = \frac{\dot{E}_\mathrm{jet}}{c} = \eta_\mathrm{jet}(j)\Mdotacc c = \Mdotjet \vjet,
\end{equation}

\noindent where $\vjet$ is the jet velocity, and $\eta_\mathrm{jet}(j)$ is the jet energy efficiency. Rearranging, we find $\psijet$ in terms of the free parameter $\vjet$,

\begin{equation}
    \label{eq:adaf_psijet}
    \psijet \equiv \eta_\mathrm{jet}(j)\bigg(\frac{c}{\vjet}\bigg).
\end{equation}

\noindent The sub-grid jet energy and momentum are

\begin{equation}
    \label{eq:adaf_subgrid_jet_energy}
    \dot{E}_\mathrm{jet,sub} = \frac{\eta_\mathrm{jet}}{1 + \psijetsub + \psiadaf}\Mdotin c^2,
\end{equation}

\noindent and

\begin{equation}
    \label{eq:adaf_subgrid_jet_momentum}
    \dot{P}_\mathrm{jet,sub} = \frac{\eta_\mathrm{jet}}{1 + \psijetsub + \psiadaf} \Mdotin c,
\end{equation}

\noindent respectively.  Given the momentum constraint, the resolved energy, momentum, and mass fluxes in the jet follow immediately as, 

\begin{equation}
    \label{eq:adaf_energy_flux_jet}
    \dot{E}_\mathrm{jet} = \frac{1}{2}\frac{\psijet}{1 + \psijetsub + \psiadaf}\Mdotin \vjet^2
\end{equation}

\begin{equation}
    \label{eq:adaf_momentum_flux_jet}
    \dot{P}_\mathrm{jet} = \frac{\psijet}{1+\psijetsub+\psiadaf} \Mdotin \vjet,
\end{equation}

\noindent and,

\begin{equation}
    \label{eq:adaf_mass_flux_jet}
    \Mdotjet = \frac{\psijet}{1 + \psijetsub + \psiadaf} \Mdotin,
\end{equation}

\noindent respectively. Our model assumes that  $\dot{P}_\mathrm{jet,sub} = \dot{P}_\mathrm{jet}$, leading to energy losses in the jet that scale with the resolved jet velocity

\begin{equation}
    \label{eq:jet_energy_ratio}
    \frac{\dot{E}_\mathrm{jet}}{\dot{E}_\mathrm{jet,sub}} = \frac{1}{2}\frac{\vjet}{c}.
\end{equation}

To compute $\psiadaf$, we assume that the hot ADAF outflows are energy conserving,

\begin{equation}
    \label{eq:adaf_wind_energy_balance}
    \frac{1}{2} \Mdotwind \vwind^2 = \epsadaf \etadisc \Mdotacc c^2,
\end{equation}

\noindent where $\epsadaf$ is a free parameter that describes the coupling of the energy to the wind, and $\vwind$ is the wind velocity free parameter. Rearranging gives us an equation for $\psiadaf$,

\begin{equation}
    \label{eq:adaf_psiadaf}
    \psiadaf \equiv 2\epsadaf\etadisc\bigg(\frac{c}{\vwind}\bigg)^2.
\end{equation}

Similar to the jet, the energy, momentum, and mass rates are fully specified as,

\begin{equation}
    \label{eq:adaf_energy_flux_wind}
    \dot{E}_\mathrm{wind} = \frac{\epsadaf\etadisc}{1+\psijetsub+\psiadaf} \Mdotin c^2,
\end{equation}

\begin{equation}
    \label{eq:adaf_momentum_flux_wind}
    \dot{P}_\mathrm{wind} = \frac{\sqrt{2\epsadaf\etadisc\psiadaf}}{1+\psijetsub+\psiadaf} \Mdotin c,
\end{equation}

\noindent and,

\begin{equation}
    \label{eq:adaf_mass_flux_wind}
    \Mdotwind = \frac{\psiadaf}{1+\psijetsub+\psiadaf} \Mdotin,
\end{equation}

\noindent respectively.

\section{Simulation methodology}
\label{sec:methods} 

To test our new black hole feedback model, we use the \textsc{Simba} sub-grid galaxy formation model as a base \citep{Dave2019}.  \textsc{Simba} has models for cooling, star formation, stellar feedback, and dust evolution that have been shown to well-reproduce observations of galaxy populations.  The model exists in the \pkg{GIZMO} code \citep{Hopkins2015a} which uses the Lagrangian mesh-free finite mass method \citep{Lanson2008a, Lanson2008b, Gaburov2011}. 

In the following sub-sections, we discuss the relevant black hole sub-grid models adapted from \pkg{Simba} that we use in our study.  For more details on the cooling, star formation, stellar feedback, and dust models, we refer the reader to \cite{Dave2019}.  In all of our simulations throughout this study, we use the same cosmological parameters as \cite{Dave2019}, which we show in Table~\ref{tbl:cosmology}.

\begin{table}
\centering
 \caption{In all of our cosmological simulations we use a cosmology consistent with the observed parameters from the \citealt{Ade2016}.}
 \label{tbl:cosmology}
 \begin{tabular}{cc}
  \hline
  Parameter & Value \\                            
  \hline
  \hline
  $\Omega_\mathrm{m}$    &
  0.3 \\
  $\Omega_\mathrm{\Lambda}$ &
  0.7 \\
  $\Omega_\mathrm{b}$ &
  0.048 \\
  $h$ &
  0.68 \\
  $\sigma_\mathrm{8}$ &
  0.82 \\
  $n_\mathrm{s}$ &
  0.97 \\
  \hline
 \end{tabular}
\end{table}

\subsection{Seeding}
\label{sec:methods_seeding}

In the \pkg{Simba} model, black holes of mass $10^4\,\Msunh$ are seeded into galaxies once they become resolved at $\sim100$ particles.  That mass corresponds to $M_\mathrm{galaxy} = 3\times10^{9}$ $\Msun$.  One behaviour of the \pkg{Simba} model is that black holes typically grow rapidly to the black hole stellar mass relationship, and then grow along that relationship as cold gas removed by the quasar mode feedback balances the accretion rate.

\subsection{Accretion and feedback}
\label{sec:methods_accretion}

We model the movement of gas into the environment of the supermassive black holes via a Bondi accretion estimator combined with an estimator based on the cold gas available within the black hole neighbourhood ($\dot{M}_\mathrm{cold}$),

\begin{equation}
    \Mdotin = \dot{M}_\mathrm{Bondi} + \dot{M}_\mathrm{cold}.
    \label{eq:accretion_inflow}
\end{equation}

\noindent To separate the two phases of gas, we use a temperature cut of $T = 10^5$ $\mathrm{K}$ where we only consider $T > 10^5$ $\mathrm{K}$ for the Bondi accretion calculations.  If gas is at temperatures $T < 10^5$ $\mathrm{K}$ and is sufficiently dense to be considered star forming ($n_\mathrm{H} > 0.13$ $\mathrm{cm}^{-3}$), we use that gas in the calculation for $\dot{M}_\mathrm{cold}$.

For Bondi accretion, we use the usual formulation

\begin{equation}
    \dot{M}_\mathrm{Bondi} = \frac{4\pi G^2 M_\mathrm{BH}^2 \rho_\mathrm{gas}}{(c_\mathrm{s}^2 + v_\mathrm{BH}^2)^{3/2}},
    \label{eq:bondi_accretion}
\end{equation}

\noindent where $M_\mathrm{BH}$ is the mass of the SMBH, $\rho_\mathrm{gas}$ is the surrounding hot gas density, $c_\mathrm{s}$ is the hot gas sound speed, and $v_\mathrm{BH}$ is the relative velocity of the SMBH with respect to the surrounding hot gas.  

For cold gas accretion, we follow \cite{Qiu2019} and use

\begin{equation}
    \dot{M}_\mathrm{cold} = \epscold\dfrac{M_\mathrm{cold}}{t_\mathrm{ff}},
    \label{eq:torque_accretion}
\end{equation}

\noindent where $M_\mathrm{cold}$ is the cold gas mass surrounding the black hole, $t_\mathrm{ff} = (3\pi / 32G\rho)^{1/2}$ is the free fall time, and $\epscold$ is a free parameter describing the efficiency of accretion.  We sum all of the mass within the neighbourhood of the black hole to compute $\rho$ in the free fall time.

Our black hole model for the radiative efficiencies is added directly into the \pkg{Simba} model, with modifications to the wind and jet mass loadings, and temperature of the jet.  For more details on the numerical implementation of the \pkg{Simba} model, see \cite{Dave2019}.  We briefly describe the original model and modifications below.  Note that everything we discuss in this section regarding our quasar mode and slim disc mode are equivalent, since our numerical implementation is identical except for the mass loadings.

First, for the quasar mode, \pkg{Simba} uses the implementation of \cite{Choi2012a} that assumes some fraction $\facc$ of the inflowing material $\Mdotin$ is driven out as a wind with mass rate $\Mdotwind$.  That fraction is the same fraction that we use in equation~\ref{eq:quasar_mdotacc},

\begin{equation}
    \label{eq:feedback_accretion_fraction}
    \facc = \frac{1}{1 + \psi},
\end{equation}

\noindent where $\psi = \Mdotwind / \Mdotin$ is the mass loading any of the modes we described in the previous section.  At a given time step $\Delta t$, we compute the accretion rate using equation~\ref{eq:accretion_inflow} and subsequently the total mass that should be accreted in that step as

\begin{equation}
    \label{eq:feedback_accreted_mass}
    M_\mathrm{acc} = \facc \Mdotin \Delta t.
\end{equation}

Recall that \pkg{Simba} has two separate masses for the black hole (BH) particles: a dynamical ($\Mbhdyn$) and a sub-grid mass ($\Mbhsub$).  The physical mass represents the dynamical mass for the gravity calculations, and the sub-grid mass represents the true, physical mass of the BH on sub-grid scales.  The separation between the masses is necessary since our particle mass resolution is much coarser than the seed mass. If the dynamical mass were equal to the seed mass, then the dynamics of the simulated BHs would be noisy, leading to BHs possibly being ejected from the cores of galaxies.  At each time-step, we advance $\Mbhsub$ using the accretion rate estimator,

\begin{equation}
    \label{eq:feedback_advance_msub}
    (\Mbhsub)_\mathrm{k+1} = (\Mbhsub)_\mathrm{k} + M_\mathrm{acc},
\end{equation}

\noindent where $(\Mbhsub)_\mathrm{k}$ is the sub-grid BH mass at time-step $k$.

In the case that $\Mbhsub > \Mbhdyn$, the physical mass has exceeded the dynamical mass and we must extract mass from surrounding particles into the BH to conserve mass.  First, we loop over all gas particles within the kernel and compute a probability for accretion,

\begin{equation}
    \label{eq:feedback_accretion_prob}
    p_\mathrm{i} = (\Mbhsub - [\Mbhdyn + M_\mathrm{acc,j
    }]) \frac{W_\mathrm{ij}}{\facc\rho_\mathrm{gas}},
\end{equation}

\noindent where $p_\mathrm{i}$ is the probability of accreting mass from particle $i$, $M_\mathrm{acc,j}$ is the cumulative mass counted for accretion in this loop for BH $j$, $W_\mathrm{ij}$ is the kernel weight between $i$ and $j$, and $\rho_\mathrm{gas}$ is the gas density surrounding BH $j$.  As ($\Mbhdyn + M_\mathrm{acc,j}) \to \Mbhsub$, $p_\mathrm{i} \to 0$, ensuring that no further mass is accreted.  We boost the probability by $\facc$ to account for feedback, as we only accrete $\facc M_\mathrm{gas,part}$ from each gas particle, and the remainder we drive as a wind.  Here $M_\mathrm{gas,part}$ is the mass of a gas particle in our simulation.  At each time step, we generate a uniformly distributed random number in the interval $w_\mathrm{i} \in [0, 1)$ and only accrete (or select to drive as a wind) from particles with $p_\mathrm{i} < w_\mathrm{i}$.

It is important to note that mass is automatically conserved when we only accrete $\facc$ of each gas particle, as we boosted our probability by $1 / \facc$.  Recall that the accretion fraction corresponds to $1/(1+\psi)$ and the outflow fraction $\psi / (1 + \psi)$, respectively.

In the case when $\Mbhsub < \Mbhdyn$, mass conservation is not necessary as the total mass accreted has not reached the resolution of the simulation.  Therefore, we only have to ensure that feedback processes are resolved,

\begin{equation}
    \label{eq:feedback_accretion_prob_unresolved}
    p_\mathrm{i} = ([1 - \facc] \Mdotin) \Delta t\, \bigg(\frac{W_\mathrm{ij}}{\rho_\mathrm{gas}}\bigg),
\end{equation}

\noindent where $(1-\facc)\Mdotin = \Mdotwind$.

Particles selected for accretion are also selected for feedback.  Once a particle is selected, we subtract the necessary mass directly from the simulation particle mass.  Note that in each BH time-step, we ensure that radiation is accounted for by subtracting off the radiative efficiencies (recall equations \ref{eq:slim_disc_mdotbh}, \ref{eq:quasar_mass_growth}, \& \ref{eq:adaf_final_mdotacc}).  Then, we launch the remaining mass at some velocity $\vwind$ or $\vjet$, depending on the mode of feedback.  We identify these particles as \textit{wind particles} and, following \cite{Dave2019}, decouple them from the hydrodynamics solver and only allow them to feel the effects of gravity.  They are decoupled for only a short period of time, and then they fully recouple to the medium.  In all cases, we decouple the wind particles for $10^{-4} t_\mathrm{H}(z)$, where $t_\mathrm{H}(z)$ is the Hubble time at a redshift $z$.  For the quasar and slim disc mode, we choose the local angular momentum direction of the gas within the kernel, $\hat{L}$, to kick the particles,

\begin{equation}
    \label{eq:feedback_kick_velocity}
    (\mvec{v}_\mathrm{gas,part})_\mathrm{k+1} = (\mvec{v}_\mathrm{gas,part})_\mathrm{k} \pm \vwind\,\hat{L},
\end{equation}

\noindent where $(\mvec{v}_\mathrm{gas,part})_\mathrm{k}$ is the gas particle velocity at time-step $k$.  There is a $50\%$ chance of alignment or anti-alignment with $\hat{L}$.  The jet follows the same procedure, except we also heat the gas particles before launch to $T = 10^{8}\,\mathrm{K}$ and in some calibrations choose a randomly selected isotropic direction.

\subsection{Dynamics}
\label{sec:methods_dynamics}

Physically, a supermassive black hole (SMBH) moving through a gravitational field experiences a friction force due to its own gravitational wake.  The dynamical friction causes SMBHs to lose energy in their orbits and slowly sink to the centre of their host galaxies.  In cosmological simulations, the gravitational resolution is often at the $\sim\mathrm{kpc}$ scale and, therefore, dynamical friction is unresolved\footnote{Although there are attempts to model sub-grid scale dynamical friction forces, see e.g. \citealt{Tremmel2017a}.}.  For that reason, we assume that SMBHs that form in our simulations are pinned in the center of their galaxy.  Specifically, at each timestep we check each galaxy for its most massive black hole and move it directly to the most bound star particle in the galaxy to ensure it does not escape due to numerical effects.

\subsection{Galaxy and Halo Finding}
\label{sec:methods_galaxy_and_halo_finding}

In all of our cosmological simulations throughout this work we need information about galaxies and their host dark matter halos.  We use the \pkg{CAESAR} package\footnote{\url{https://caesar.readthedocs.io/}} that uses a friends-of-friends algorithm to find galaxies via groups of cold gas and stars.  It also uses this algorithm to find dark matter halos, and then links the two together into a single galaxy-host relationship.  

\section{Calibration}
\label{sec:calibration}

Our aim is to provide a more physically motivated model for black hole feedback in cosmological simulations.  For that reason, we calibrate our model parameters in cosmological simulations that contain thousands of galaxies.  We choose a box size of $L = 25 \cMpch$ with $N_\mathrm{gas} = 256^3$ and $N_\mathrm{dark} = 256^3$.  This gives a mass resolution of $M_\mathrm{gas} \approx 1.24\times10^{7} \Msunh$ for gas and $M_\mathrm{dark} \approx 6.51\times10^{7} \Msunh$ for dark matter.

Our new black hole feedback model introduces several new parameters, and we must determine their values.  However, computational resources limit our abilities to search all of the available parameter space and we must fix a large fraction of the parameters in order to explore the model further.  In total we vary $6$ parameters in our model, giving us a total of 576 simulations.  First, we explain the fixed parameters and the motivation for those values.

\begin{table}
\centering
 \caption{The $6$ parameters that we vary for our calibration. The jet can either be energy or momentum conserving on the large scales. $\epsadaf$ is the coupling of the secondary feedback component in the ADAF mode. $\Madaf$ is the black hole mass at which the ADAF mode is allowed. $\vjet$ is the jet velocity.  $j$ is the spin of all black holes.  We also test the importance of the black hole jet direction by allowing it to be isotropic, or non-isotropic.}
 \label{tbl:parameter_variations}
 \begin{tabular}{ccccc}
  \hline
  Parameter & Value 1 & Value 2 & Value 3 & Value 4 \\                            
  \hline
  \hline
  Jet Loading                  & Momentum          & Energy           & -  & -      \\
  $\epsadaf$                & 0.001          & 0.01          & 0.1  & -       \\
  $\Madaf$ ($\Msun$)    & $0$ & $3\times10^7$ & $7\times10^7$ & $10^8$   \\
  $\vjet$ ($\kms$)      & $5000$             & $7500$            & $10^4$ & $1.5\times10^4$          \\
  $j$                       & 0.3           & 0.6          & 0.9 & -    \\
  Isotropic Jet             & Yes           & No            & - & -        \\
  \hline
  \hline
  $\epssd$                  & 0.05          & -             & -         \\
  $\epsq$                   & 0.05          & -             & -         \\
  $\vwindsd$                & $1000\,\kms$  & -             & -         \\
  $\vwindq$                 & $1000\,\kms$  & -             & -         \\
  $\lowbound$        & 0.03          & -             & -         \\
  $\upbound$        & 0.3           & -             & -         \\
  \hline
 \end{tabular}
\end{table}

\subsection{Fixed parameters}
\label{sec:calibration_fixed}

There are two fixed parameters that describe the coupling of the energy from the black hole (BH) feedback in our model: $\epssd$ and $\epsq$.  The former is for the slim disc mode (high accretion rates) and the latter is for the quasar mode (mid-range accretion rates).  Both of these parameters could be independent, but to reduce computational costs we fix both values to $\epssd = \epsq = 0.05$.  Note that the coupling factor for the momentum driven feedback is degenerate with the cold gas accretion efficiency, $\epscold$.  The feedback coupling sets the normalisation of the BH-stellar mass relationship by lowering the cold gas fraction in the BH environment.  Simultaneously, $\epscold$ determines how much of that gas may be used to grow the BH.  Therefore, we chose $\epsq$ and $\epssd$ to be low values since $\epscold$ must be less than unity (i.e., it is an efficiency).  Using lower values of $\epssd$ and $\epsq$ actually result in a much lower mass loading than $\sim 20 L / c$ but we choose to express everything in this form to separate the numerics from the physics.

\begin{figure}
    \centering
    \includegraphics[scale=0.52]{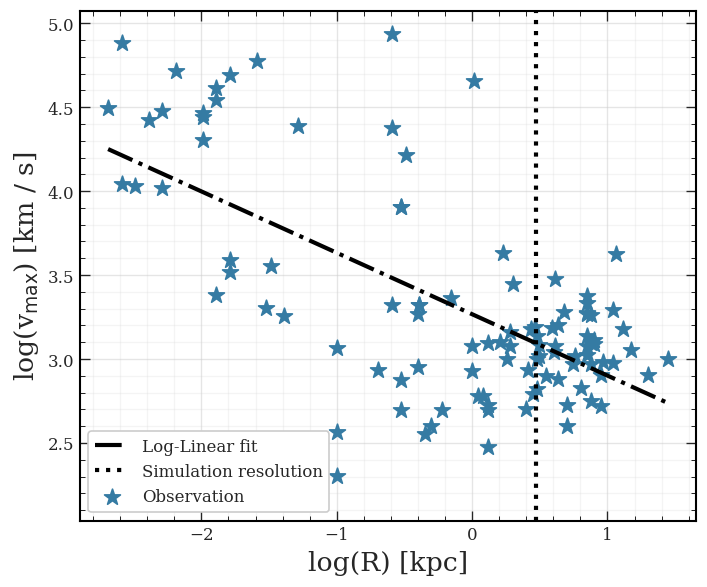}
    \caption{Active galactic nuclei wind velocities as a function of radius produced from \citealt{Fiore2017}.  We fit a power law for illustrative purposes and found a relationship $\mathrm
{log}(v_\mathrm{max}) = -0.366\,\mathrm{log}(R) + 3.267$.  For a typical cosmological simulation with resolution $\sim2\,\kpch$, wind velocities should be of order $\sim1000\,\kms$.}
    \label{fig:agn_wind_fiore}
\end{figure}

Next, the wind velocities $\vwindq$ and $\vwindsd$ must be fixed in order to determine the accretion fractions in both the quasar and slim disc feedback modes.  Similarly to the coupling factors, we fix $\vwind = \vwindq = \vwindsd$ due to the lack of observations of wind speeds from rapidly accreting black holes.  The observed physical wind velocities from quasar-like AGN range from $\sim 10^2$ $\kms$ to $\sim 10^5$ $\kms$ depending on the observed scale.  We compiled wind maximum velocity results as a function of radius from \cite{Fiore2017} in Fig.~\ref{fig:agn_wind_fiore}.  The dotted vertical line shows the simulation gravitational softening of $\sim 2 \kpch$.  For \textit{illustrative purposes only}, we show a log-linear fit to the observed data.  Evidently, by the time winds launched from close to the black holes approach the scale of our simulation resolution, they should have already slowed down to  $\sim 10^3$ $\kms$.  For that reason, we fix $\vwind = 10^3$ $\kms$.  Again, we must emphasise that the wind velocity is degenerate with $\epscold$, and the feedback efficiency parameters.  We attempt to use physically motivated values when possible to remove degeneracy.

The last two fixed parameters are the boundary accretion rate ratios separating the three modes: $\lowbound$ and $\upbound$.  These accretion ratios determine when there is a break in the accretion efficiency, $\eta = \eta(j, \Mdotacc)$.  We follow \cite{Merloni2008}  and set $\lowbound = 0.03$ (see also \citealt{Maccarone2003, Greene2006}) so that below this value, $\eta \propto \Mdotacc / \MdotEdd$.  For the high accretion rate regime, we set $\upbound = 0.3$, as above this value the classical thin-$\alpha$-disc model is suggested to break down \citep{Laor1989, McClintock2006, Sadowski2009, Madau2014a}.

\subsection{Varied parameters}
\label{sec:calibration_varied}

The 6 parameters that we choose to vary are: the jet loading, $\epsadaf$, $\Madaf$, $\vjet$, $j$, and whether the jet is isotropic or aligned with the angular momentum of the galaxy. All of the varied parameters (see Table~\ref{tbl:parameter_variations}), except for $j$, only impact the low accretion rate regime, i.e. $\Mdotacc / \MdotEdd < \lowbound$, in our calibration tests.  As discussed in Section~\ref{sec:new_model_slim_disc}, $j$ impacts the high accretion rate regime significantly through the accretion fraction $\facc$.  All of these parameters either do not have strong constraints or do not have a physical meaning, but only exist due to the insufficient resolution of our simulations.

\subsection{Observational constraints and parameter choice}
\label{sec:calibration_constraints}

One of the more common methods for calibrating cosmological simulations is to compare some observational constraints at $z = 0$ to the simulated data, representing many variations of the free parameters, at the same redshift.  Our choices in Sections~\ref{sec:calibration_fixed} \& \ref{sec:calibration_varied} lead to 576 parameter variations that must be constrained.  That is a significant number of simulations that would need to be run to $z = 0$ and so we tried an unconventional approach of calibrating our parameter variations to observations at $z = 2$ (corresponding to $33\%$ of the run-time, compared to $z = 0$) -- allowing us to remove a large part of parameter space for parameter variations that have already failed by this time.  We only ran the simulations that gave the best match to observations past $z = 2$, according to our criteria described below.

\begin{table*}
\centering
 \caption{The $6$ parameters that we vary for our calibration and their values for calibration simulations that were accepted under our constraints on the galaxy stellar mass function, black hole-stellar mass relation, and quenched density at $z = 2$.}
 \label{tbl:parameter_variations_accepted}
 \begin{tabular}{ccccccc}
  \hline
  Run label & $\epsadaf$ & $\Madaf$ ($\Msun$) & Loading Type & $\vjet$ & $j$ & Isotropic Jet \\   
  \hline
  \hline
$008$    &    0.001    &    0    &   Energy    &    5000    &   0.3    &    No    \\
$041$    &    0.001    &    $5\times10^7$    &   Energy    &    5000    &   0.3    &    Yes    \\
$058$    &    0.001    &    $10^8$    &   Energy    &    7500    &   0.3    &    No    \\
$059$    &    0.001    &    $10^8$    &   Energy    &    7500    &   0.3    &    Yes    \\
$061$    &    0.001    &    $10^8$    &   Energy    &    10000    &   0.3    &    Yes    \\
$063$    &    0.001    &    $10^8$    &   Energy    &    15000    &   0.3    &    Yes    \\
$130$    &    0.001    &    0    &   Momentum    &    7500    &   0.9    &    No    \\
$131$    &    0.001    &    0    &   Momentum    &    7500    &   0.9    &    Yes    \\
$147$    &    0.001    &    $10^7$    &   Momentum    &    7500    &   0.9    &    Yes    \\
$161$    &    0.001    &    $5\times10^7$    &   Momentum    &    5000    &   0.9    &    Yes    \\
$163$    &    0.001    &    $5\times10^7$    &   Momentum    &    7500    &   0.9    &    Yes    \\
$\mathbf{165}$    &    \textbf{0.001}    &    $\mathbf{5\times10^7}$    &   \textbf{Momentum}    &    \textbf{10000}    &   \textbf{0.9}    &    \textbf{Yes}    \\
$179$    &    0.001    &    $10^8$    &   Momentum    &    7500    &   0.9    &    Yes    \\
$181$    &    0.001    &    $10^8$    &   Momentum    &    10000    &   0.9    &    Yes    \\
$183$    &    0.001    &    $10^8$    &   Momentum    &    15000    &   0.9    &    Yes    \\
$218$    &    0.010    &    $10^7$    &   Energy    &    7500    &   0.3    &    No    \\
$222$    &    0.010    &    $10^7$    &   Energy    &    15000    &   0.3    &    No    \\
$223$    &    0.010    &    $10^7$    &   Energy    &    15000    &   0.3    &    Yes    \\
$224$    &    0.010    &    $5\times10^7$    &   Momentum    &    5000    &   0.3    &    No    \\
$232$    &    0.010    &    $5\times10^7$    &   Energy    &    5000    &   0.3    &    No    \\
$233$    &    0.010    &    $5\times10^7$    &   Energy    &    5000    &   0.3    &    Yes    \\
$235$    &    0.010    &    $5\times10^7$    &   Energy    &    7500    &   0.3    &    Yes    \\
$238$    &    0.010    &    $5\times10^7$    &   Energy    &    15000    &   0.3    &    No    \\
$253$    &    0.010    &    $10^8$    &   Energy    &    10000    &   0.3    &    Yes    \\
$254$    &    0.010    &    $10^8$    &   Energy    &    15000    &   0.3    &    No    \\
$299$    &    0.010    &    $5\times10^7$    &   Energy    &    7500    &   0.6    &    Yes    \\
$313$    &    0.010    &    $10^8$    &   Energy    &    5000    &   0.6    &    Yes    \\
$385$    &    0.100    &    0    &   Momentum    &    5000    &   0.3    &    Yes    \\
$387$    &    0.100    &    0    &   Momentum    &    7500    &   0.3    &    Yes    \\
$389$    &    0.100    &    0    &   Momentum    &    10000    &   0.3    &    Yes    \\
$391$    &    0.100    &    0    &   Momentum    &    15000    &   0.3    &    Yes    \\
$399$    &    0.100    &    0    &   Energy    &    15000    &   0.3    &    Yes    \\
$414$    &    0.100    &    $10^7$    &   Energy    &    15000    &   0.3    &    No    \\
$417$    &    0.100    &    $5\times10^7$    &   Momentum    &    5000    &   0.3    &    Yes    \\
$419$    &    0.100    &    $5\times10^7$    &   Momentum    &    7500    &   0.3    &    Yes    \\
$421$    &    0.100    &    $5\times10^7$    &   Momentum    &    10000    &   0.3    &    Yes    \\
$423$    &    0.100    &    $5\times10^7$    &   Momentum    &    15000    &   0.3    &    Yes    \\
$425$    &    0.100    &    $5\times10^7$    &   Energy    &    5000    &   0.3    &    Yes    \\
$429$    &    0.100    &    $5\times10^7$    &   Energy    &    10000    &   0.3    &    Yes    \\
$432$    &    0.100    &    $10^8$    &   Momentum    &    5000    &   0.3    &    No    \\
$434$    &    0.100    &    $10^8$    &   Momentum    &    7500    &   0.3    &    No    \\
$436$    &    0.100    &    $10^8$    &   Momentum    &    10000    &   0.3    &    No    \\
$438$    &    0.100    &    $10^8$    &   Momentum    &    15000    &   0.3    &    No    \\
$442$    &    0.100    &    $10^8$    &   Energy    &    7500    &   0.3    &    No    \\
$444$    &    0.100    &    $10^8$    &   Energy    &    10000    &   0.3    &    No    \\
$446$    &    0.100    &    $10^8$    &   Energy    &    15000    &   0.3    &    No    \\
$452$    &    0.100    &    0    &   Momentum    &    10000    &   0.6    &    No    \\
$472$    &    0.100    &    $10^7$    &   Energy    &    5000    &   0.6    &    No    \\
$492$    &    0.100    &    $5\times10^7$    &   Energy    &    10000    &   0.6    &    No    \\
$493$    &    0.100    &    $5\times10^7$    &   Energy    &    10000    &   0.6    &    Yes    \\
$507$    &    0.100    &    $10^8$    &   Energy    &    7500    &   0.6    &    Yes    \\
$510$    &    0.100    &    $10^8$    &   Energy    &    15000    &   0.6    &    No    \\
$569$    &    0.100    &    $10^8$    &   Energy    &    5000    &   0.9    &    Yes    \\
$572$    &    0.100    &    $10^8$    &   Energy    &    10000    &   0.9    &    No    \\
$573$    &    0.100    &    $10^8$    &   Energy    &    10000    &   0.9    &    Yes    \\
$575$    &    0.100    &    $10^8$    &   Energy    &    15000    &   0.9    &    Yes    \\
  \hline
 \end{tabular}
\end{table*}

We choose three observational constraints\footnote{See \cite{Schaye2014} for an excellent discussion of why galaxy stellar masses and black hole (BH) masses cannot be predicted in cosmological simulations.} for our parameter variations: (i) the galaxy stellar mass function (GSMF), (ii) the BH mass-stellar mass correlation, and (iii) the quenched galaxy density.   While constraints (i) and (ii) are common, constraint (iii) is also necessary because \pkg{Simba}-like models (such as ours) will have difficulties producing sufficient quenched galaxies at high-redshift.

Our BH model is \pkg{Simba}-like insofar as we decided to use our radiative efficiency curve in a specific implementation of BH feedback.  As previously discussed, our BH feedback model is truly the radiative efficiency curves in Fig.~\ref{fig:agn_power_eta} and could be attached to any contemporary BH feedback implementation in Lagrangian or Eulerian based cosmological simulations.  However, the kinetic jet and kinetic quasar wind in \pkg{Simba} (which uses a Lagrangian hydrodynamical solver) under-predicts high-redshift quenched galaxy densities \citep{Merlin2019}.  That is in the nature of the model itself: the quasar mode does not have sufficient energy to quench galaxies, while the kinetic jet only activates after $z \sim 3$.  The latter is either a \textit{feature} of the model or a consequence of calibration.  Regardless, we choose to calibrate our simulations to the quenched galaxy density at $z = 2$ in attempt to alleviate the issue.

Our $(25 \cMpch)^3$ volumes are small and, therefore, may not contain a sufficient sample size to predict the correct quenched density.  Therefore, we do not place a strong constraint on the quenched number density, but rather a minimum bound.  The results of \cite{Merlin2019} (their Fig. $7$) suggest that there are somewhere between $\approx10^{-4}$ to $5\times10^{-4}$ quenched galaxies per $\Mpccubed$ at $z = 2$.  That gives approximately $5$ to $15$ quenched galaxies in our simulated volume.  We are only really concerned about removing large parts of parameter space by stopping at $z = 2$, therefore we set the constraint that \textit{there must be at least one}\footnote{We simulated the same initial conditions with \pkg{Simba}, and found zero quenched galaxies at $z = 2$.} quenched galaxy in our volume to proceed the simulation to $z = 0$. 

Constraints (i) and (ii) also do not require strict\footnote{Or, \textit{hard} constraints in the language of constraint programming.} constraints since we are simply removing large sections of parameter space.  For (i), we require the simulated GSMFs to have a galaxy number density $\lesssim 10^{-4}$ $\invMpccubed$ at $\Mstar \sim 3\times10^{11}\,\Msun$, following the observations in \cite{Tomczak2014}.  For (ii), we produced the BH mass-stellar mass relationship by binning our simulated BH masses in bins of galaxy stellar masses (width $\Delta \log(M_*) = 0.2$) with the constraint that the galaxies are well-resolved, i.e. $M_* \gtrsim 10^{10}\,\Msun$.  We then found the mean value and standard deviation in each bin.   We impose the constraint that the fit for the BH mass-stellar mass relationship in \cite{Kormendy2013} must lay within one standard deviation of the simulated mean value curve to be acceptable.  However, we found that using the accretion rate in equation~\ref{eq:torque_accretion} with a value of $\epscold = 0.06$ rarely leads to deviations from the \cite{Kormendy2013} fit.  Note that observations at $z = 2$ suggest that BHs may be above the \cite{Kormendy2013} relationship.  However, we know from experience with the \pkg{Simba} model that (a) BH feedback will be ineffective if the BHs are below the relationship, (b) BH feedback will be over-effective if the BHs are above the relationship, and, more importantly, (c) the normalisation of the BH-stellar mass relationship is invariant in \pkg{Simba}-like models.  Since it is invariant, if the BH masses are much greater than the \cite{Kormendy2013} relationship at high redshift, they will remain above for all cosmic time and galaxies will be over-quenched.  Therefore, we must calibrate the relationship at $z = 2$, knowing they will stay on the relationship at $z = 0$.

Out of all of our 576 calibration simulations, only 56 satisfied all three of our constraints at $z = 2$.  Interestingly, the quenched density was the dominant constraint that removed the bulk of parameter space, indicating that it is indeed difficult for \pkg{Simba}-like models to quench high-redshift galaxies.  We show the parameters that gave us reasonable results in Table~\ref{tbl:parameter_variations_accepted}.  It is immediately obvious that there seem to be no clearly favoured values for any of the parameters.  These 56 runs span the possible values for $\epsadaf$, and have no preference for the loading type, or $\vjet$.  However, there is a preference for $\Madaf > 0$, as well as lower values of $j \sim 0.3$.  Additionally, there is a marginal preference for having an isotropic jet.

It is important to note that the quenched density constraint may be responsible for the limited number of successful runs not only because it is strict, but also due to the stochastic nature of the simulations themselves.  Recall that we define quenched galaxies as those with $\mathrm{sSFR} < 10^{-11}\,\mathrm{yr}^{-1}$ and that there are few massive galaxies within the small volumes we simulated.  When we evaluated the $\mathrm{sSFR}$-$M_*$ relationship, we noticed that in some cases there were galaxies which were quite close to being quenched, but did not make the threshold.  The inherent stochasticity combined with a firm constraint led to many calibration values being rejected.  We again stress that the radiative efficiency model we present in this work does not necessarily have to be combined with a \pkg{Simba}-like model, and could be combined into any black hole feedback scheme, leading to a different calibration.  We only discuss these 56 calibrations and their parameter values as a proof-of-concept.

\begin{figure*}
    \centering
    \includegraphics[scale=0.4]{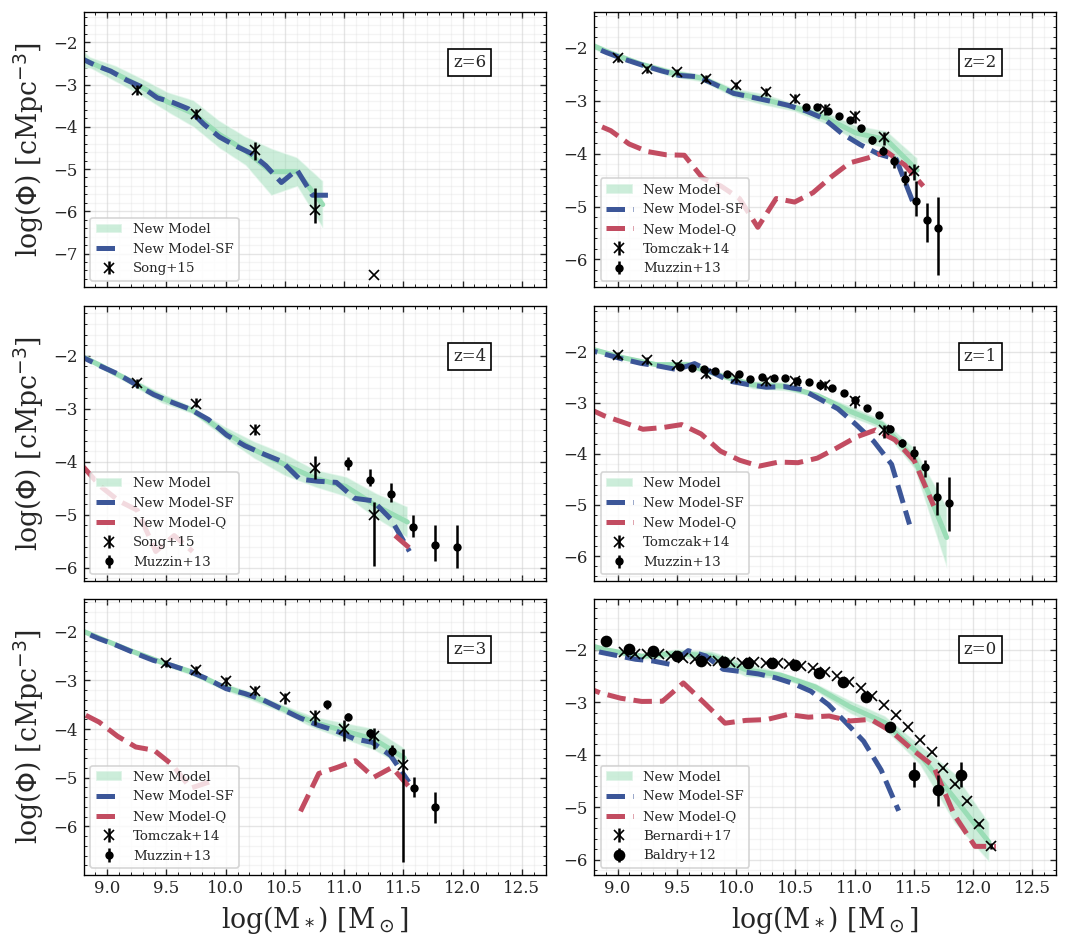}
    \caption{The observational and simulated galaxy stellar mass functions at $z = 6$, $4$, $3$, $2$, $1$, and $z = 0$ from top-left to bottom-right, respectively.  We use observations from \citealt{Song2016} for $z = 6$ and $z = 4$, \citealt{Muzzin2013} and \citealt{Tomczak2014} for $z = 3, 2$ and $z = 1$, and \citealt{Baldry2012} and \citealt{Bernardi2017} for $z = 0$.  We use calibration run 165 from Table~\ref{tbl:parameter_variations_accepted} in a $(100\,\cMpch)^3$ volume, labeled as \textit{New Model}, as a cyan shaded region (showing cosmic variance).  We further divide the volume into both quenched and star forming galaxies at an $\mathrm{sSFR}$ cut of $10^{-11}$ $\mathrm{yr}^{-1}$.  The quenched galaxies are shown as a red dashed line, and the star forming galaxies as a blue dashed line.  Our new model provides a reasonable match to the galaxy stellar mass function across all cosmic epochs, only under-predicting the knee of the galaxy stellar mass function by a factor of $\sim 2$.}
    \label{fig:galaxy_stellar_mass_function_fiducial}
\end{figure*}

Not only are these calibrations able to reproduce the GSMF and the BH mass-stellar mass relationship but they are also able to broadly reproduce the observed cosmic star formation rate density (CSFRD) of the Universe down to $z = 0$, even though they were only calibrated to the three constraints at $z = 2$.  While it is promising that the model calibrations reproduce a single unconstrained observable, we require verification in a larger volume to ensure better statistics and probe larger galaxy mass ranges, since the CSFRD is very sensitive to box size.

\section{Simulated galaxy population survey}
\label{sec:galaxies}

It is important to determine how well our new black hole (BH) feedback model can reproduce a realistic galaxy population.  However, our calibration volume contained few galaxies on the mass scale where BH feedback is expected to be the most impactful -- massive galaxies and galaxy groups/clusters.  A much larger volume is necessary to probe these mass scales and, therefore, we decided to simulate a larger volume using a single parameter choice -- our run labelled 165 (see Table~\ref{tbl:parameter_variations_accepted}).  We chose this particular run since it had the lowest mean-squared error in the galaxy stellar mass function (GSMF), compared to observations, at $z = 0$.  Our simulation consists of one run with $N_\mathrm{gas} = N_\mathrm{dark} = 1024^3$ in a $(100 \cMpch)^3$ volume.  The large-volume simulation uses exactly the same sub-grid models and cosmology as described in Section~\ref{sec:methods}, and have the same particle mass resolution as our calibration suite described at the beginning of Section~\ref{sec:calibration}.

\subsection{Calibration constraints revisited}
\label{sec:galaxies_constraints_revisited}

In Fig.~\ref{fig:galaxy_stellar_mass_function_fiducial} we show the simulated galaxy stellar mass function (GSMF) at $6$ redshifts from our large $(100\,\cMpch)^3$ volume.  The redshifts are $z = 6, 4, 3$ in the left column from top to bottom, respectively, and then $z = 2, 1, 0$ from top to bottom in the right column, respectively.  For redshifts $z \geqslant 4$, we compare with the observations in \cite{Song2016}, whereas for $1 \leqslant z \leqslant 3$ we compare with both observational results from \cite{Muzzin2013} and \cite{Tomczak2014}.  For $z = 0$, we compare with both \cite{Baldry2012} and \cite{Bernardi2017}.  At each redshift we separate quenched (red) and star forming (blue) galaxies to show their contribution to the overall GSMF (shaded region).  The overall GSMF includes a shaded region showing the impact of cosmic variance.  At redshifts $z \geqslant 3$ the match to the GSMF is mostly due to star forming galaxies and is not due to the BH feedback model that we introduced.  Rather, it is a result of the \pkg{Simba} cooling, star formation, and stellar feedback models providing a good fit to the observations.  At redshifts $z \lesssim 4$, the BH feedback model begins to impact the quenched fraction within the volume, as is evidenced by the bottom left panels, where it is clear that massive galaxies are beginning to quench.  By $z = 1$, the quenched population dominates the overall GSMF at the high-mass end, providing a reasonable match to the observations.  However, at $z = 0$, our BH model provides a reasonable fit to the \cite{Bernardi2017} observations, where the entire contribution to the overall GSMF above $M_* \sim 10^{11}\,\Msun$ is due to the quenched population.  However, the model under-predicts the abundance of galaxies at the knee of the GSMF due to over-quenching.

\begin{figure}
    \centering
    \includegraphics[scale=0.52]{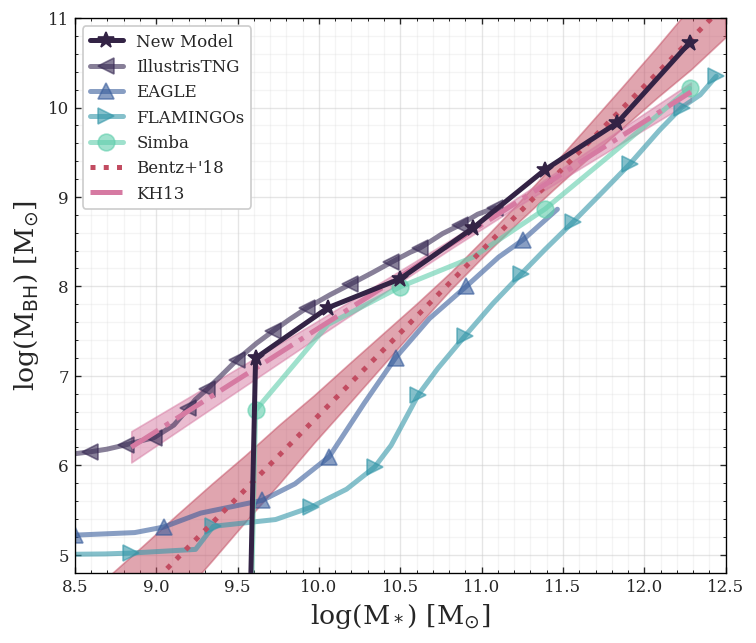}
    \caption{The black hole mass-stellar mass relationship.  We show our new model in the $(100\,\cMpch)^3$ volume as a black solid line with star markers, in bins of $M_*$.  We compare other large-volume simulations with markers given in the legend.  The shaded regions show the \citealt{Kormendy2013} observed fit (KH13; dash-dotted) and the \citealt{Bentz2018} observed fit (dotted).  In resolved galaxies ($M_* > 3\times10^9$ M$_\odot$), our new model matches the normalisation and slope of the KH13 relationship.}
    \label{fig:galaxy_bh_stellar_mass}
\end{figure}

Fig.~\ref{fig:galaxy_bh_stellar_mass} shows the BH mass-stellar mass relation at $z = 0$ for our new model along with comparisons to \pkg{IllustrisTNG}, \pkg{EAGLE}, \pkg{FLAMINGO}, and \pkg{Simba}.  We include observational relations from \cite{Kormendy2013} as well as \cite{Bentz2018} as a comparison.  Our model provides an reasonable fit for $M_* > 10^{10}\,\Msun$ at $z = 0$, and effectively follows the results from the \pkg{Simba} simulation.  This indicates that the simpler cold gas accretion model we use in this work is just as effective as the more complicated torque accretion model presented in \cite{Angles-Alcazar2017a} and used in the \pkg{Simba} simulations.  The resolution of a galaxy in our simulation ($\sim100$ particles) is $\sim 3\times10^9\,\Msun$ and, therefore, the fit diverges at the low mass end.  We emphasise that we calibrated to the relationship in \cite{Kormendy2013} and only show the \cite{Bentz2018} relationship as it demonstrates the variety in observational results where other (e.g. \pkg{EAGLE} and \pkg{FLAMINGO}) simulations may provide a reasonable match to the slope.

Overall, the results of our calibration method used in this work are promising.  The GSMF at $z = 0$ is in reasonable agreement with observations, giving us confidence that our model is viable and that we can trust the stellar mass build-up in our simulated galaxy population.  Similarly, our simulated BH-mass stellar mass relationship at $z = 0$ shows that our cold accretion model provides a reasonable growth rate with no divergences from the \cite{Kormendy2013} relationship.  The latter point is important and shows how, fundamentally, BH mass growing with the stellar mass build up could be more easily explained by the presence of cold gas in the cores of the galaxies\footnote{At least in these coarse grained models. }.

\begin{figure}
    \centering
    \includegraphics[scale=0.52]{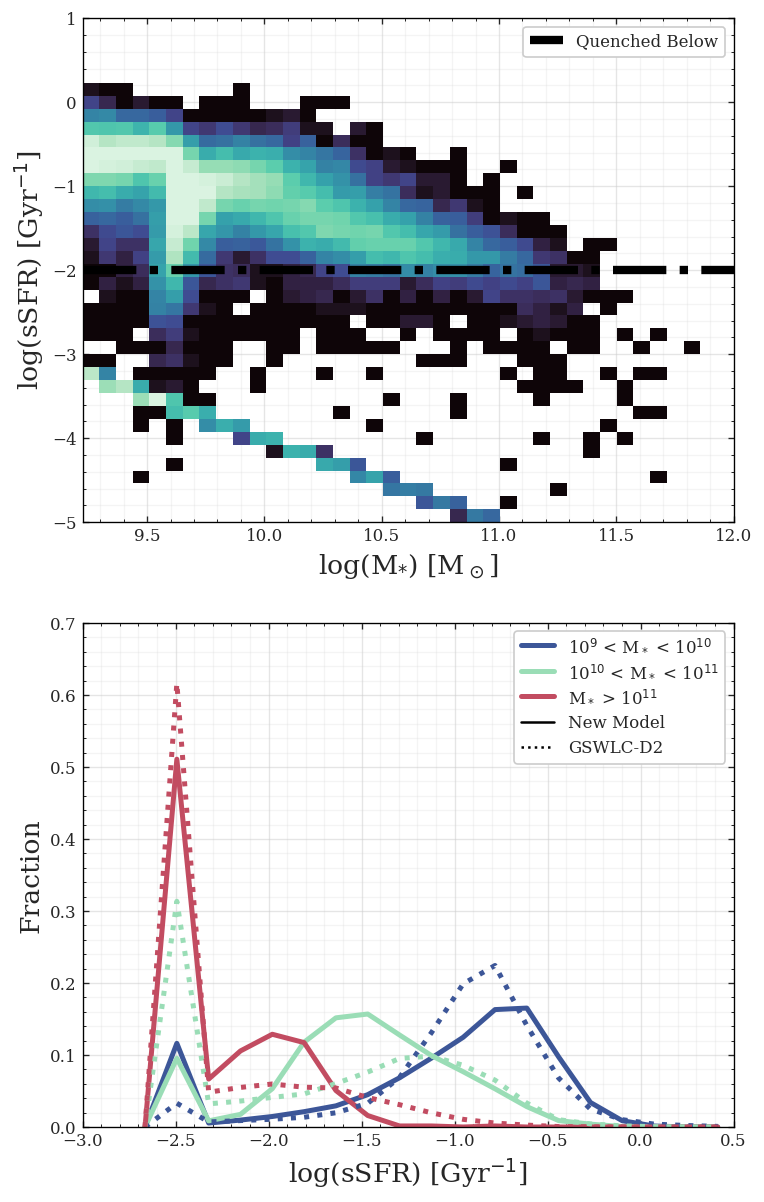}
    \caption{(top) A two-dimensional histogram showing the log-normalised number of galaxies from our large-volume calibrated simulation, at $z = 0$, in bins of specific star formation rate ($\mathrm{sSFR}$) and stellar mass ($M_*$).  The dash-dotted line shows the upper limit in $\mathrm{sSFR}$ for considering a galaxy quenched.  Our new model produces a star forming main sequence as well as a quenched distribution.  (bottom) The one-dimensional $\mathrm{sSFR}$ distribution from our calibrated large-volume simulation at $z = 0$, with each bin normalised to the total number of galaxies in the simulation.  We divide the galaxies into three mass bins, as shown in the legend (solid curves).  For comparison, we show observations from \pkg{GSWLC-D2}.  Our new model over predicts the number of quenched galaxies in the low-mass bin, as well as the high-mass bin.}
    \label{fig:galaxy_ssfr}
\end{figure}

\subsection{Star formation rates and efficiencies}
\label{sec:galaxies_sfrs}

The fundamental explanation for observed massive quenched galaxies in the Universe is that black hole (BH) feedback is sufficiently powerful to halt star formation in these systems -- both by local injection of energy and by heating and lowering the density of gas in the halo, balancing cooling.  In the previous Section, we demonstrate that our BH feedback model prevents the build up of stellar mass in galaxies --- reproducing an exponential tail in the galaxy stellar mass function.  In this Section, we investigate the specific star formation rates (sSFRs) of our large-volume galaxy population in order to determine if the number of quenched systems matches the observations.

The top panel in Fig.~\ref{fig:galaxy_ssfr} shows a two-dimensional distribution of sSFRs (in $\mathrm{Gyr}^{-1}$) and stellar masses, with the brightness of each bin indicating the log-normalised count of galaxies in that bin.  The dash-dotted line shows the observationally-motivated separation between quenched galaxies ($\mathrm{sSFR} \lesssim 10^{-2}$ $\mathrm{Gyr}^{-1}$) and star forming galaxies ($\mathrm{sSFR} \gtrsim 10^{-2}$ $\mathrm{Gyr}^{-1}$).  The power-law overdensity of galaxies in the lower left section of the panel are galaxies with $\mathrm{SFR} = 0$ $\Msun\,\mathrm{yr}^{-1}$.  Our simulated galaxy population has a separate quenched population towards the lower right of this distribution \citep{Bell2004}.  We do not include the observational results from SDSS here for clarity, but we now discuss the collapsed distribution of sSFRs.

The bottom panel in Fig.~\ref{fig:galaxy_ssfr} shows the one-dimensional sSFR distribution of our simulated sample (solid lines), showing the fraction of galaxies in each bin.  We additionally divide the population into three bins of stellar mass: (i) $10^{9} < \Mstar / \Msun < 10^{10}$ (blue), (ii) $10^{10} < \Mstar / \Msun < 10^{11}$ (green), and (iii) $\Mstar > 10^{11}\,\Msun$ (red).  The dotted lines show the observational results from \pkg{GALEX-SDSS-WISE} catalog, DR2 \citep{Salim2016, Salim2018}.  

In our low mass galaxies, there is a slight shift in the location of the peak toward higher $\mathrm{sSFR}$ values and there are $\sim10\%$ quenched galaxies versus $\sim4\%$ expected from observations.  In the middle mass range, effectively the knee of the GSMF, the distribution is peaked, but offset from the observations. The model under-predicts the fraction of quenched galaxies in this range by 20\%.  In the high mass range, our model slightly over-predicts the number of star forming galaxies while providing a reasonable match to the number of quenched galaxies.  The star forming massive galaxies have a broader distribution of observed $\mathrm{sSFR}$s than our model.

\begin{figure}
    \centering
    \includegraphics[scale=0.52]{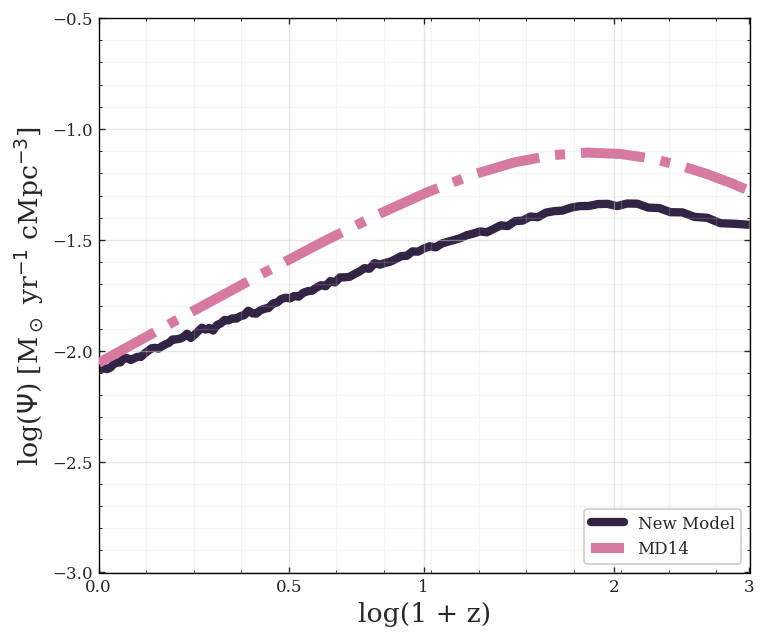}
    \caption{The star formation rate density history for our large-volume calibrated simulation from $z = 3$ to $z = 0$ (solid curve).  The observational fit from \citealt{Madau2014} is shown as a dash-dotted curve (MD14).  Our new model, although calibrated to completely different observables at $z = 0$ and $z = 2$ and in a much smaller volume, provides a reasonable prediction to the slope of the relationship from the peak at $z \approx 2$ to $z = 0$.}
    \label{fig:csfh_big_volume}
\end{figure}

In Fig.~\ref{fig:csfh_big_volume}, we show the cosmic star formation rate density (CSFRD) history from our large-volume simulation.  The dash-dotted line shows the observation compilation from \cite{Madau2014}.  We do not show z > 3 since the CSFRD is modulated by the stellar feedback.  While the approximate shape is predicted correctly, the normalisation is a factor of $1.5\times$ below the peak.  This again signals that the model is over-quenching low-mass galaxies, and aligns with the GSMF.

\begin{figure}
    \centering
    \includegraphics[scale=0.52]{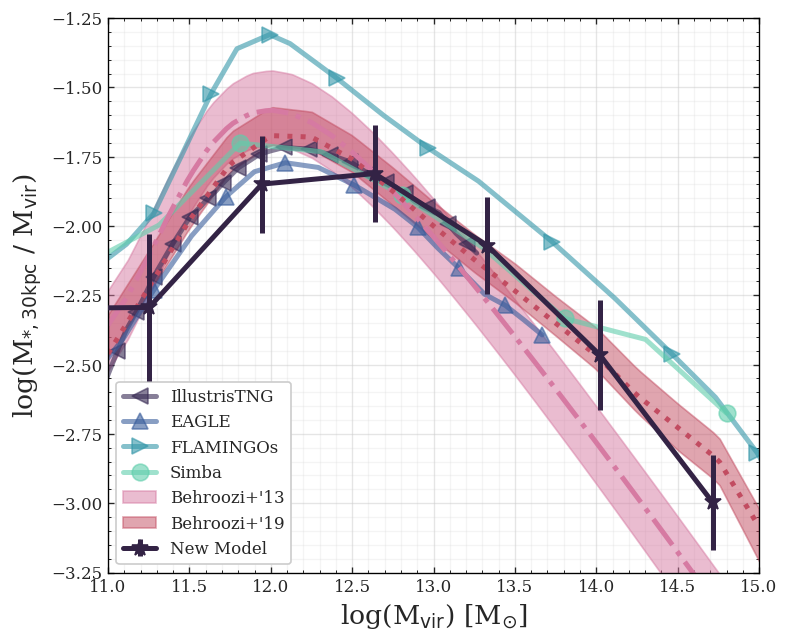}
    \caption{The stellar mass-halo mass relationship at $z = 0$, for central galaxies only.  We show abundance matching results from \citealt{Behroozi2013} and \citealt{Behroozi2019} as the shaded regions with dash-dotted and dotted lines, respectively.  Our new model is the solid curve with star markers, and error bars given by the standard deviation of the value in each bin of $\Mvir$. For comparison we show results taken from \pkg{IllustrisTNG}, \pkg{EAGLE}, \pkg{FLAMINGO}, and \pkg{Simba}.  Our model predicts a reasonable match to the abundance matching results, and especially succeeds at suppressing star formation in the brightest cluster galaxies at the galaxy-cluster scale.}
    \label{fig:galaxy_stellar_efficiency}
\end{figure}

In Fig.~\ref{fig:galaxy_stellar_efficiency} we show the stellar mass-halo mass relationship (SMHM) for our simulated population, indicated by the line with a star symbol and error bars showing a standard deviation about the mean value in each bin of stellar mass.  We show results from \pkg{IllustrisTNG} (left triangle), \pkg{EAGLE} (upwards triangle), \pkg{FLAMINGO} (right triangle), and \pkg{Simba} (circle) for comparison.  Instead of showing the observations directly, we show the abundance matching results from both \cite{Behroozi2013} and \cite{Behroozi2019} for clarity.  The shaded regions indicate their provided uncertainty bounds for their model calculations.  Notice that in the updated abundance matching model, the high-mass end slope has become shallower.  \pkg{IllustrisTNG}, \pkg{EAGLE}, and \pkg{Simba} provide the best predictions for the SMHM relationship, while \pkg{FLAMINGO}  diverges from all other models above roughly $\Mvir\gtrsim10^{12}\,\Msun$.  

Our model provides a reasonable fit at masses $\Mvir \gtrsim 10^{12}\,\Msun$, but begins to diverge near $\Mvir \lesssim 10^{12}\,\Msun$.  In particular, our model is successful at predicting a good estimate of the stellar efficiencies in the group and cluster regime, which we define as halos in the mass range $\Mvir \gtrsim 10^{13}\,\Msun$.  Unfortunately, our model under-predicts the stellar efficiencies in lower mass halos compared to the updated \cite{Behroozi2019} model.  The disagreement is only at a $0.25 \,\mathrm{dex}$ level (less than a factor of approximately $2$).  In the group and cluster regime, our BH feedback model sufficiently prevents stellar mass build up across cosmic time that broadly agrees with the abundance matching predictions.

\subsection{Gas fractions}
\label{sec:galaxies_gas_fractions}

\begin{figure}
    \centering
    \includegraphics[scale=0.52]{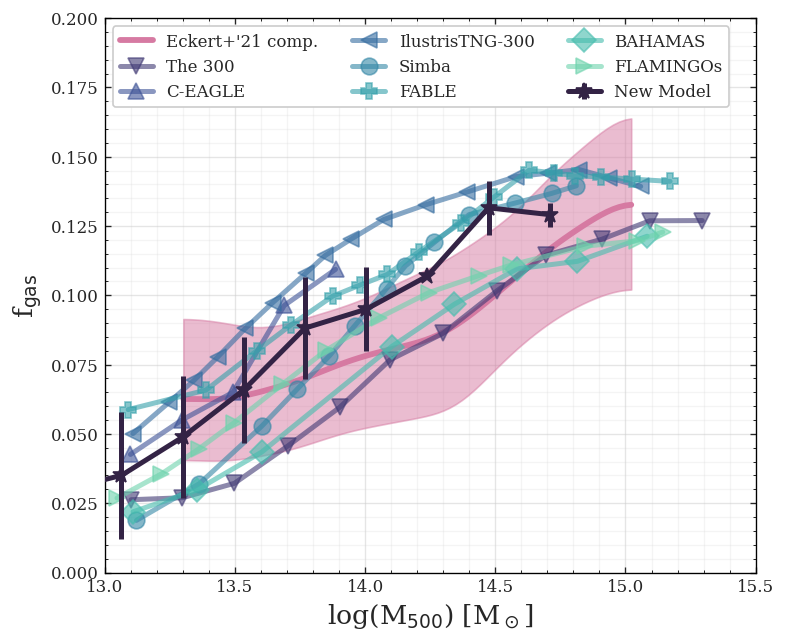}
    \caption{The hot gas fractions within galaxy groups and clusters.  The shaded region with the solid pink line shows the observational compilation from the review in \citealt{Eckert2021}.  We show our new model in black with star markers.  For comparison, we show simulation results from \pkg{The 300}, \pkg{C-EAGLE}, \pkg{IllustrisTNG-300}, \pkg{Simba}, \pkg{FABLE}, \pkg{BAHAMAS}, and \pkg{FLAMINGO}, taken from Fig. 15 in the review paper.  Our new model tracks the \pkg{IllustrisTNG-300} model, while showing a promising trend toward the observed gas fractions in more massive clusters.}
    \label{fig:galaxy_m500_gas_fracs}
\end{figure}

One important observational constraint for black hole (BH) feedback models is the total baryon fraction within galaxy groups and clusters.  The universal ratio of baryons to total matter in the Universe depends on the cosmology as $f_\mathrm{baryon} = \Omegab / \Omegam$, which for our selected cosmology (see Section~\ref{sec:methods}) gives $f_\mathrm{baryon} = 0.16$.  Observationally, only the most massive halos, i.e. galaxy clusters, retain this fraction within their virial radii (cf. Fig. 8 in \citealt{Liang2016}), leading to the idea that BH feedback is responsible for pushing gas outside of the virial radii of halos with masses $\Mvir \lesssim 10^{14}\,\Msun$.  The timing of gas ejection from the halos is unclear in theoretical models as it is still an active area of research (see \citealt{Eckert2021}; \citealt{Oppenheimer2021}).  It is important to note that the baryon fractions within the cores of halos, i.e. $\lesssim \Rfivehundred$, are difficult to constrain as they rely on assumptions of hydrostatic equilibrium and occasionally extrapolation since it is always not possible to measure out to $\Rfivehundred$.  Despite these difficulties, it is clear that there is a lack of baryons within the cores of galaxy groups and low-mass clusters, and any successful BH feedback model should be able to reproduce this trend.

Fig.~\ref{fig:galaxy_m500_gas_fracs} shows the hot gas ($T > 3\times10^5$ $\mathrm{K}$) fractions $M_\mathrm{gas} / M_\mathrm{total}$ within $\Rfivehundred$ of the center of each halo.   Our model is the solid line with stars.  We show simulation results from the 300 collaboration (downward triangle), \pkg{C-EAGLE} (upward triangle), \pkg{IllustrisTNG-300} (left triangle), \pkg{Simba} (circle), \pkg{FABLE} (plus sign), \pkg{BAHAMAS} (diamond), and \pkg{FLAMINGO} (right triangle) for comparison.  \cite{Eckert2021} compiled observations into a single trend line with error, which we show as a solid line and shaded region, respectively.  

Our new model produces gas fractions in line with those observed at the low-mass end of the group scale, jumping up at $M_\mathrm{500} \sim 3\times10^{13}\,\Msun$.  At higher masses, up into the cluster scale, the gas fractions match the predictions from \pkg{FABLE}.  Note that we did not calibrate to this relationship, did not test this mass scale, or investigate this redshift in our calibration simulations -- it is a prediction of our model. Interestingly, our gas fractions remain above the observations while our stellar efficiencies (i.e., $M_* / M_\mathrm{vir}$) remain low.  That is counter-intuitive to most of the BH feedback models available, as decreasing the gas fraction is paramount to effectively quenching the central brightest cluster galaxy.  As an example, \pkg{FLAMINGO} was calibrated to the gas fractions in Fig.\ref{fig:galaxy_m500_gas_fracs}, but overall has much higher stellar efficiencies than any other simulation.  That indicates that perhaps our model is able to allow more control over the gas fractions while still being able to provide a quenched galaxy population.

Overall, the majority of simulations predict a slope that is too steep at all masses --- indicating that the BH feedback models predict all galaxy groups and clusters have effectively a similar baryon fraction, counter to the observations.  Notable exceptions are the 300 collaboration, \pkg{C-EAGLE}, and \pkg{BAHAMAS} simulations, which provide reasonable predictions for the normalisation and the slope of the relationship.  A full investigation into the impact of the model on the group and cluster scale is beyond the scope of this study, but we will investigate group and cluster baryon fractions in much more detail in a follow-up study.

\section{Conclusions}
\label{sec:conclusions}

The exponential decline in the observed number density of galaxies at high stellar masses suggests that powerful black hole (BH) feedback is a necessary component in our theoretical understanding of galaxy evolution.  Models of BH feedback in cosmological simulations generally rely heavily on modelling feedback via radiation powered by accretion onto the BH, $E_\mathrm{feedback} = \eta\Mdotacc c^2$.  Typically, studies assume a quasar-like mode with a constant, spin-independent radiative efficiency $\eta \sim 0.1$, occasionally combined with a more powerful jet-like mode at low accretion rates to quench massive galaxies.  We introduce the \textsc{Obsidian} sub-grid BH model: a model for large-scale cosmological simulations motivated by small-scale calculations of individual accretion flows that combines three physical regimes dependent on the accretion rate onto the BH, $\eta = \eta(j, \Mdotacc)$.  We divided the three regimes into the low accretion rate regime $\Mdotacc / \MdotEdd \lesssim 0.03$, middle regime $0.03 \lesssim \Mdotacc / \MdotEdd \lesssim 0.3$, and high regime $\Mdotacc / \MdotEdd \gtrsim 0.3$ following an approach similar to the semi-analytic work of \cite{Merloni2008}.

In the low accretion rate regime, we model BH accretion and feedback as an advection dominated accretion flow (ADAF) combined with a large-scale jet.  In an ADAF, gas densities are low, such that much of the energy advects into the BH before it can radiate, leading to a low radiative efficiency.  On the scales of cosmological simulations ($\sim\mathrm{kpc}$) this model allows for two simultaneous feedback components: an isotropic outflow term combined with a collimated jet.  The isotropic outflow term has a radiative efficiency that scales with the accretion rate as $\eta(j, \Mdotacc) \propto \Mdotacc / \MdotEdd$ with the normalisation set by continuity across the entire $\Mdotacc / \MdotEdd$ range for $\eta$.  We model the jet component following the calculations in \cite{Benson2009} as a disc wind+jet, with disc wind efficiency taken from their Fig. 3.  We follow \cite{Talbot2020} and \cite{Husko2022} and implement a Blandford-Znajek \citep{Blandford1977} jet.

For the middle accretion rate regime, or the quasar mode, we use a constant radiative efficiency $\eta = \eta(j)$ motivated by a model of a geometrically thin accretion disc.  The normalisation of $\eta$ in this case comes from the simulations in \cite{Sadowski2009}, allowing us to set the normalisation in the low accretion rate regime as well.  

In the high accretion rate regime, we adopt the optically-thick, geometrically-slim disc model.  At these accretion rates, energy dissipated in the disc is moved radially inward faster than it can radiate, leading to a lower radiative efficiency with increasing $\Mdotacc$.  We used the radiative efficiency fits from the simulations of \cite{Sadowski2009} that lead to a behaviour $\eta(j, \Mdotacc) \propto (\Mdotacc / \MdotEdd)^{-1}$.  

While the radiative efficiency function we provide in Section~\ref{sec:new_model} could be applied to any implementation of BH feedback in cosmological simulations, we choose to incorporate it in the \pkg{Simba} galaxy formation model.  \pkg{Simba} provides models for cooling, star formation, and stellar feedback over which we could test the impact of BH feedback on the galaxy population.  We inherit the BH seeding and repositioning algorithm from \pkg{Simba}, but made a slight alteration to the accretion rate estimator.  While \pkg{Simba} uses the torque accretion mode from \cite{Hopkins2011} and \cite{Angles-Alcazar2017a}, we introduce a cold accretion rate that scales with the cold gas mass near the BH as $\Mdotin \propto M_\mathrm{cold} / t_\mathrm{ff}$, where $t_\mathrm{ff}$ is the free-fall time estimated at the maximum influence radius of the BH.  We found a proportionality constant of $\epscold = 0.06$ reproduced the behaviour of the more complicated torque accretion model, although it is degenerate with the quasar mode feedback normalisation.

Using our modified \pkg{Simba} model as a base, we run a calibration suite of 576 cosmological simulations that varied 6 of our model parameters including the directionality of the jet, i.e. either isotropic or collimated , and whether it was energy or momentum loaded.  We use observations and physically-motivated arguments to fix the other $6$ parameters of our model.  Our 576 cosmological simulations consisted of volumes $(25\cMpch)^3$ with particle mass resolutions of $\Mgas \approx 1.24\times10^{7}\,\Msunh$ and $M_\mathrm{dark} \approx 6.51\times10^7\,\Msunh$ which were sufficiently large to contain a few galaxy-group scale objects.  To constrain our parameters, we used the galaxy stellar mass function \citep{Tomczak2014}, BH mass-stellar mass relationship \citep{Kormendy2013}, and the quenched number density of galaxies \citep{Merlin2019} --- all at a redshift of $z = 2$. 

Out of the hundreds of calibration simulations, only 56 were viable to continue to $z = 0$.  The main constraint that removes a significant portion of parameter space was the quenched galaxy number density as it seems difficult to quench galaxies at higher redshift in our model as implemented into a \pkg{Simba}-like framework.  Given the low number of viable solutions, we use the calibration that had the lowest mean squared error compared to the observed galaxy stellar mass function at $z = 0$ and $z = 2$ as our fiducial model, but in a significantly larger volume.

We simulate a $(150\,\cMpc)^3$ volume to $z = 0$ with our fiducial model with $1024^3$ gas and $1024^3$ dark matter particles, which gave the same resolution as our calibration tests.  In this volume, we investigate the galaxy stellar mass function, BH mass-stellar mass relation, specific star formation rate distribution, stellar mass-halo mass relation, and the baryon fractions within $\Rfivehundred$ of galaxy group-scale objects.

First, we analyse our three calibration constraints.  The galaxy stellar mass functions across all redshifts provide reasonable fits.  The BH mass-stellar mass relationship at also matched the results from \cite{Kormendy2013}.  Our model unfortunately predicts too many quenched lower mass galaxies ($\Mstar\lesssim10^{11}\,\Msun$), but provides a good estimation of the number of quenched, massive galaxies ($\Mstar\gtrsim10^{11}\,\Msun$).

Second, we analyse how well our model extrapolates to unconstrained observables: the stellar mass-halo mass function and the baryon fraction within galaxy groups and clusters.  

Our model predicts a slightly lower normalisation in the stellar mass-halo mass relation at masses $\Mvir \lesssim 3\times10^{12}\,\Msun$ but overall matches the normalisation and slope of the abundance matching results from \cite{Behroozi2019}.  The peak of the relationship matched well, as well as the efficiencies at the galaxy cluster scale.  These slight differences could be calibrated out of the model, by using the stellar mass-halo mass relationship as an additional constraint, if required.

The hot gas fraction within $\Rfivehundred$ is a useful observational constraint for BH feedback models as this is the region most impacted from the powerful outflows.  In Fig.~\ref{fig:galaxy_m500_gas_fracs}, we showed the hot gas fraction $f_\mathrm{gas}$ as a function of $\Mfivehundred$ for group to cluster scale halos ($\Mfivehundred\gtrsim 10^{13}\,\Msun$) halos in our large-volume simulation.  We found that our model provides a reasonable match to the observations in \cite{Eckert2021}.  Interestingly, our model provides a reasonably quenched galaxy population with halos that have higher than observed gas fractions.  That implies that gas is not being effectively removed from the halo and that the energy from the BH is injected more locally.

Overall, the \textsc{Obsidian} sub-grid model generates a suitable galaxy population for study. The main drawback is the overabundance of quenched, lower mass galaxies ($\Mstar\lesssim3\times10^{10}\,\Msun$).  While we explore a wide range of possible parameters in this work, it is indeed possible that we have not found the set of parameters that best represents the galaxies across all mass scales.  In this work, we focused heavily on the ADAF mode since it contains most of the new features when compared with \pkg{Simba} but computational resources limits how many parameters may be varied at once.  One important step forward would be to design a parameter study for all parameters related to the slim disk and quasar modes, since those modes should most impact the lower mass galaxies.  We additionally encourage further development of the model by adding a self-consistent spin distribution for the BHs, which could open up many research avenues into the impact of BHs on their galactic environs.

\section*{Acknowledgements}

This research was enabled in part by support provided by the Digital Research Alliance of Canada.  The majority of the simulations in this research were made possible by SciNet and the Niagara supercomputing cluster.  DR and AB acknowledges the support of the Natural Sciences and Engineering Research Council of Canada (NSERC; funding reference number 534263) and through the Discovery Grant program.  DR is supported by the Simons Foundation.  AB acknowledges support from the Infosys Foundation via an endowed Infosys Visiting Chair Professorship at the Indian Institute of Science.  Some of the simulations presented in this work were run on the Flatiron Institute's research computing facilities (the Iron compute cluster), supported by the Simons Foundation.  The authours thank Daniel Angl\'{e}s-Alc\'{a}zar, Sophie Koudmani, Fred Jennings, Alessandro Lupi, Christopher C. Hayward, Megan Tillman, and Asya Rennehan for useful advice during the course of this research.  Additionally, we thank Dominique Eckert, and Annalisa Pillepich for providing observational and simulation data used for comparison in this work.  Finally, we thank the anonymous referee for helping to improve this work.

Our analysis was performed using the Python programming language.  The following packages were used throughout the analysis: \pkg{h5py} \citep{Collette2013}, \pkg{numpy} \citep{Harris2020}, \pkg{scipy} \citep{Virtanen2020}, \pkg{yt} \citep{Turk2011}, and \pkg{matplotlib} \citep{Hunter2007}.  Prototyping of the analysis scripts was performed in the IPython environment \citep{Perez2007}.

\section*{Data availability}

The data underlying this article will be shared on reasonable request to the corresponding authour.  We encourage further community development of the model, and will provide the source code on request.




\bibliographystyle{mnras}
\bibliography{main} 

\begin{thebibliography}{}
\makeatletter
\relax
\def\mn@urlcharsother{\let\do\@makeother \do\$\do\&\do\#\do\^\do\_\do\%\do\~}
\def\mn@doi{\begingroup\mn@urlcharsother \@ifnextchar [ {\mn@doi@} {\mn@doi@[]}}
\def\mn@doi@[#1]#2{\def\@tempa{#1}\ifx\@tempa\@empty \href {http://dx.doi.org/#2} {doi:#2}\else \href {http://dx.doi.org/#2} {#1}\fi \endgroup}
\def\mn@eprint#1#2{\mn@eprint@#1:#2::\@nil}
\def\mn@eprint@arXiv#1{\href {http://arxiv.org/abs/#1} {{\tt arXiv:#1}}}
\def\mn@eprint@dblp#1{\href {http://dblp.uni-trier.de/rec/bibtex/#1.xml} {dblp:#1}}
\def\mn@eprint@#1:#2:#3:#4\@nil{\def\@tempa {#1}\def\@tempb {#2}\def\@tempc {#3}\ifx \@tempc \@empty \let \@tempc \@tempb \let \@tempb \@tempa \fi \ifx \@tempb \@empty \def\@tempb {arXiv}\fi \@ifundefined {mn@eprint@\@tempb}{\@tempb:\@tempc}{\expandafter \expandafter \csname mn@eprint@\@tempb\endcsname \expandafter{\@tempc}}}

\bibitem[\protect\citeauthoryear{Abramowicz, Czerny, Lasota  \& Szuszkiewicz}{Abramowicz et~al.}{1988}]{Abramowicz1988}
Abramowicz M.~A.,  Czerny B.,  Lasota J.~P.,   Szuszkiewicz E.,  1988, \mn@doi [The Astrophysical Journal] {10.1086/166683}, 332, 646

\bibitem[\protect\citeauthoryear{Angl{\'{e}}s-Alc{\'{a}}zar, Dav{\'{e}}, Faucher-Gigu{\`{e}}re, {\"{O}}zel  \& Hopkins}{Angl{\'{e}}s-Alc{\'{a}}zar et~al.}{2017}]{Angles-Alcazar2017a}
Angl{\'{e}}s-Alc{\'{a}}zar D.,  Dav{\'{e}} R.,  Faucher-Gigu{\`{e}}re C.-A.,  {\"{O}}zel F.,   Hopkins P.~F.,  2017, \mn@doi [Monthly Notices of the Royal Astronomical Society] {10.1093/mnras/stw2565}, 2853, 13

\bibitem[\protect\citeauthoryear{Angl{\'{e}}s-Alc{\'{a}}zar et~al.,}{Angl{\'{e}}s-Alc{\'{a}}zar et~al.}{2021}]{AnglesAlcazar2021}
Angl{\'{e}}s-Alc{\'{a}}zar D.,  et~al., 2021, \mn@doi [The Astrophysical Journal] {10.3847/1538-4357/ac09e8}, 917, 53

\bibitem[\protect\citeauthoryear{Babul \& White}{Babul \& White}{1991}]{Babul1991}
Babul A.,  White S. D.~M.,  1991, \mn@doi [Monthly Notices of the Royal Astronomical Society] {10.1093/mnras/253.1.31P}, 253, 31P

\bibitem[\protect\citeauthoryear{Babul, Sharma  \& Reynolds}{Babul et~al.}{2013}]{Babul2013}
Babul A.,  Sharma P.,   Reynolds C.~S.,  2013, \mn@doi [Astrophysical Journal] {10.1088/0004-637X/768/1/11}, 768

\bibitem[\protect\citeauthoryear{Baldry et~al.,}{Baldry et~al.}{2012}]{Baldry2012}
Baldry I.~K.,  et~al., 2012, \mn@doi [Monthly Notices of the Royal Astronomical Society] {10.1111/j.1365-2966.2012.20340.x}, 421, 621

\bibitem[\protect\citeauthoryear{Bardeen}{Bardeen}{1970}]{Bardeen1970}
Bardeen J.~M.,  1970, \mn@doi [Nature] {10.1038/226064a0}, 226, 64

\bibitem[\protect\citeauthoryear{Bardeen \& Petterson}{Bardeen \& Petterson}{1975}]{Bardeen1975}
Bardeen J.~M.,  Petterson J.~A.,  1975, \mn@doi [The Astrophysical Journal] {10.1086/181711}, 195, L65

\bibitem[\protect\citeauthoryear{Barnes et~al.,}{Barnes et~al.}{2017}]{Barnes2017}
Barnes D.~J.,  et~al., 2017, \mn@doi [Monthly Notices of the Royal Astronomical Society] {10.1093/mnras/stx1647}, 471, 1088

\bibitem[\protect\citeauthoryear{Beckmann et~al.,}{Beckmann et~al.}{2019}]{Beckmann2019}
Beckmann R.~S.,  et~al., 2019, \mn@doi [Astronomy and Astrophysics] {10.1051/0004-6361/201936188}, 631, 1

\bibitem[\protect\citeauthoryear{Beckmann, Dubois, Pellissier, Polles  \& Olivares}{Beckmann et~al.}{2022}]{Beckmann2022}
Beckmann R.~S.,  Dubois Y.,  Pellissier A.,  Polles F.~L.,   Olivares V.,  2022, \mn@doi [Astronomy & Astrophysics] {10.1051/0004-6361/202243873}, 666, A71

\bibitem[\protect\citeauthoryear{Behroozi, Wechsler  \& Wu}{Behroozi et~al.}{2013}]{Behroozi2013}
Behroozi P.~S.,  Wechsler R.~H.,   Wu H.-Y.,  2013, \mn@doi [The Astrophysical Journal] {10.1088/0004-637X/762/2/109}, 762, 109

\bibitem[\protect\citeauthoryear{Behroozi, Wechsler, Hearin  \& Conroy}{Behroozi et~al.}{2019}]{Behroozi2019}
Behroozi P.,  Wechsler R.~H.,  Hearin A.~P.,   Conroy C.,  2019, \mn@doi [Monthly Notices of the Royal Astronomical Society] {10.1093/mnras/stz1182}, 488, 3143

\bibitem[\protect\citeauthoryear{Bell et~al.,}{Bell et~al.}{2004}]{Bell2004}
Bell E.~F.,  et~al., 2004, \mn@doi [The Astrophysical Journal] {10.1086/420778}, 608, 752

\bibitem[\protect\citeauthoryear{Benson \& Babul}{Benson \& Babul}{2009}]{Benson2009}
Benson A.~J.,  Babul A.,  2009, \mn@doi [Monthly Notices of the Royal Astronomical Society] {10.1111/j.1365-2966.2009.15087.x}, 397, 1302

\bibitem[\protect\citeauthoryear{Bentz \& Manne-Nicholas}{Bentz \& Manne-Nicholas}{2018}]{Bentz2018}
Bentz M.~C.,  Manne-Nicholas E.,  2018, \mn@doi [The Astrophysical Journal] {10.3847/1538-4357/aad808}, 864, 146

\bibitem[\protect\citeauthoryear{Bernardi, Meert, Sheth, Fischer, Huertas-Company, Maraston, Shankar  \& Vikram}{Bernardi et~al.}{2017}]{Bernardi2017}
Bernardi M.,  Meert A.,  Sheth R.~K.,  Fischer J.-L.,  Huertas-Company M.,  Maraston C.,  Shankar F.,   Vikram V.,  2017, \mn@doi [Monthly Notices of the Royal Astronomical Society] {10.1093/mnras/stx176}, 467, stx176

\bibitem[\protect\citeauthoryear{Berti \& Volonteri}{Berti \& Volonteri}{2008}]{Berti2008}
Berti E.,  Volonteri M.,  2008, \mn@doi [The Astrophysical Journal] {10.1086/590379}, 684, 822

\bibitem[\protect\citeauthoryear{Blandford \& Znajek}{Blandford \& Znajek}{1977}]{Blandford1977}
Blandford R.~D.,  Znajek R.~L.,  1977, \mn@doi [Monthly Notices of the Royal Astronomical Society] {10.1093/mnras/179.3.433}, 179, 433

\bibitem[\protect\citeauthoryear{Blanton, Randall, Clarke, Sarazin, Mcnamara, Douglass  \& Mcdonald}{Blanton et~al.}{2011}]{Blanton2011}
Blanton E.~L.,  Randall S.~W.,  Clarke T.~E.,  Sarazin C.~L.,  Mcnamara B.~R.,  Douglass E.~M.,   Mcdonald M.,  2011, \mn@doi [Astrophysical Journal] {10.1088/0004-637X/737/2/99}, 737

\bibitem[\protect\citeauthoryear{Boehringer, Voges, Fabian, Edge  \& Neumann}{Boehringer et~al.}{1993}]{Boehringer1993}
Boehringer H.,  Voges W.,  Fabian A.~C.,  Edge A.~C.,   Neumann D.~M.,  1993, \mn@doi [Monthly Notices of the Royal Astronomical Society] {10.1093/mnras/264.1.L25}, 264, L25

\bibitem[\protect\citeauthoryear{Booth \& Schaye}{Booth \& Schaye}{2009}]{Booth2009}
Booth C.~M.,  Schaye J.,  2009, \mn@doi [Monthly Notices of the Royal Astronomical Society] {10.1111/j.1365-2966.2009.15043.x}, 398, 53

\bibitem[\protect\citeauthoryear{Bourne \& Sijacki}{Bourne \& Sijacki}{2021}]{Bourne2021}
Bourne M.~A.,  Sijacki D.,  2021, \mn@doi [Monthly Notices of the Royal Astronomical Society] {10.1093/mnras/stab1662}, 506, 488

\bibitem[\protect\citeauthoryear{Bourne \& Yang}{Bourne \& Yang}{2023}]{Bourne2023}
Bourne M.~A.,  Yang H.-Y.~K.,  2023, \mn@doi [Galaxies] {10.3390/galaxies11030073}, 11, 73

\bibitem[\protect\citeauthoryear{Brienza et~al.,}{Brienza et~al.}{2023}]{Brienza2023}
Brienza M.,  et~al., 2023, \mn@doi [Astronomy & Astrophysics] {10.1051/0004-6361/202245247}, 672, A179

\bibitem[\protect\citeauthoryear{Cattaneo et~al.,}{Cattaneo et~al.}{2009}]{Cattaneo2009}
Cattaneo A.,  et~al., 2009, \mn@doi [Nature] {10.1038/nature08135}, 460, 213

\bibitem[\protect\citeauthoryear{Cenci, Sala, Lupi, Capelo  \& Dotti}{Cenci et~al.}{2020}]{Cenci2020}
Cenci E.,  Sala L.,  Lupi A.,  Capelo P.~R.,   Dotti M.,  2020, \mn@doi [Monthly Notices of the Royal Astronomical Society] {10.1093/mnras/staa3449}, 500, 3719

\bibitem[\protect\citeauthoryear{Chavan, Dabhade  \& Saikia}{Chavan et~al.}{2023}]{Chavan2023}
Chavan K.,  Dabhade P.,   Saikia D.~J.,  2023, \mn@doi [Monthly Notices of the Royal Astronomical Society: Letters] {10.1093/mnrasl/slad100}, 525, L87

\bibitem[\protect\citeauthoryear{Choi, Ostriker, Naab  \& Johansson}{Choi et~al.}{2012}]{Choi2012a}
Choi E.,  Ostriker J.~P.,  Naab T.,   Johansson P.~H.,  2012, \mn@doi [The Astrophysical Journal] {10.1088/0004-637X/754/2/125}, 754, 125

\bibitem[\protect\citeauthoryear{Choi, Ostriker, Naab, Oser  \& Moster}{Choi et~al.}{2015}]{Choi2015a}
Choi E.,  Ostriker J.~P.,  Naab T.,  Oser L.,   Moster B.~P.,  2015, \mn@doi [Monthly Notices of the Royal Astronomical Society] {10.1093/mnras/stv575}, 449, 4105

\bibitem[\protect\citeauthoryear{Choi, Ostriker, Naab, Somerville, Hirschmann, N{\'{u}}{\~{n}}ez, Hu  \& Oser}{Choi et~al.}{2017}]{Choi2017}
Choi E.,  Ostriker J.~P.,  Naab T.,  Somerville R.~S.,  Hirschmann M.,  N{\'{u}}{\~{n}}ez A.,  Hu C.-Y.,   Oser L.,  2017, \mn@doi [The Astrophysical Journal] {10.3847/1538-4357/aa7849}, 844, 31

\bibitem[\protect\citeauthoryear{Cielo, Babul, Antonuccio-Delogu, Silk  \& Volonteri}{Cielo et~al.}{2018}]{Cielo2018}
Cielo S.,  Babul A.,  Antonuccio-Delogu V.,  Silk J.,   Volonteri M.,  2018, \mn@doi [Astronomy & Astrophysics] {10.1051/0004-6361/201832582}, 617, A58

\bibitem[\protect\citeauthoryear{Ciotti, Ostriker  \& Proga}{Ciotti et~al.}{2009}]{Ciotti2009}
Ciotti L.,  Ostriker J.~P.,   Proga D.,  2009, \mn@doi [The Astrophysical Journal] {10.1088/0004-637X/699/1/89}, 699, 89

\bibitem[\protect\citeauthoryear{Collette}{Collette}{2013}]{Collette2013}
Collette A.,  2013, {Python and HDF5}.
O'Reilly

\bibitem[\protect\citeauthoryear{Conroy \& Ostriker}{Conroy \& Ostriker}{2008}]{Conroy2008}
Conroy C.,  Ostriker J.~P.,  2008, \mn@doi [The Astrophysical Journal] {10.1086/587861}, 681, 151

\bibitem[\protect\citeauthoryear{Crain \& van~de Voort}{Crain \& van~de Voort}{2023}]{Crain2023}
Crain R.~A.,  van~de Voort F.,  2023, \mn@doi [Annual Review of Astronomy and Astrophysics] {10.1146/annurev-astro-041923-043618}, 61, 473

\bibitem[\protect\citeauthoryear{Cui et~al.,}{Cui et~al.}{2022}]{Cui2022}
Cui W.,  et~al., 2022, \mn@doi [Monthly Notices of the Royal Astronomical Society] {10.1093/mnras/stac1402}, 514, 977

\bibitem[\protect\citeauthoryear{Czerny}{Czerny}{2019}]{Czerny2019}
Czerny B.,  2019, \mn@doi [Universe] {10.3390/universe5050131}, 5, 131

\bibitem[\protect\citeauthoryear{{Dalla Vecchia} \& Schaye}{{Dalla Vecchia} \& Schaye}{2012}]{DallaVecchia2012}
{Dalla Vecchia} C.,  Schaye J.,  2012, \mn@doi [Monthly Notices of the Royal Astronomical Society] {10.1111/j.1365-2966.2012.21704.x}, 426, 140

\bibitem[\protect\citeauthoryear{Dav{\'{e}}, Angl{\'{e}}s-Alc{\'{a}}zar, Narayanan, Li, Rafieferantsoa  \& Appleby}{Dav{\'{e}} et~al.}{2019}]{Dave2019}
Dav{\'{e}} R.,  Angl{\'{e}}s-Alc{\'{a}}zar D.,  Narayanan D.,  Li Q.,  Rafieferantsoa M.~H.,   Appleby S.,  2019, \mn@doi [Monthly Notices of the Royal Astronomical Society] {10.1093/mnras/stz937}, 486, 2827

\bibitem[\protect\citeauthoryear{{Di Matteo}, Springel  \& Hernquist}{{Di Matteo} et~al.}{2005}]{DiMatteo2005}
{Di Matteo} T.,  Springel V.,   Hernquist L.,  2005, \mn@doi [Nature] {10.1038/nature03335}, 433, 604

\bibitem[\protect\citeauthoryear{{Di Matteo}, Colberg, Springel, Hernquist  \& Sijacki}{{Di Matteo} et~al.}{2008}]{DiMatteo2007a}
{Di Matteo} T.,  Colberg J.,  Springel V.,  Hernquist L.,   Sijacki D.,  2008, \mn@doi [The Astrophysical Journal] {10.1086/524921}, 676, 33

\bibitem[\protect\citeauthoryear{Ding et~al.,}{Ding et~al.}{2020}]{Ding2020}
Ding X.,  et~al., 2020, \mn@doi [The Astrophysical Journal] {10.3847/1538-4357/ab5b90}, 888, 37

\bibitem[\protect\citeauthoryear{Dubois, Devriendt, Slyz  \& Teyssier}{Dubois et~al.}{2012}]{Dubois2012}
Dubois Y.,  Devriendt J.,  Slyz A.,   Teyssier R.,  2012, \mn@doi [Monthly Notices of the Royal Astronomical Society] {10.1111/j.1365-2966.2011.20236.x}, 420, 2662

\bibitem[\protect\citeauthoryear{Dubois et~al.,}{Dubois et~al.}{2014}]{Dubois2014}
Dubois Y.,  et~al., 2014, \mn@doi [Monthly Notices of the Royal Astronomical Society] {10.1093/mnras/stu1227}, 444, 1453

\bibitem[\protect\citeauthoryear{Dubois et~al.,}{Dubois et~al.}{2021}]{Dubois2021}
Dubois Y.,  et~al., 2021, \mn@doi [Astronomy & Astrophysics] {10.1051/0004-6361/202039429}, 651, A109

\bibitem[\protect\citeauthoryear{Eckert, Gaspari, Gastaldello, {Le Brun}  \& O'Sullivan}{Eckert et~al.}{2021}]{Eckert2021}
Eckert D.,  Gaspari M.,  Gastaldello F.,  {Le Brun} A. M.~C.,   O'Sullivan E.,  2021, \mn@doi [Universe] {10.3390/universe7050142}, 7, 142

\bibitem[\protect\citeauthoryear{Ellison, Teimoorinia, Rosario  \& Mendel}{Ellison et~al.}{2016}]{Ellison2016}
Ellison S.~L.,  Teimoorinia H.,  Rosario D.~J.,   Mendel J.~T.,  2016, \mn@doi [Monthly Notices of the Royal Astronomical Society: Letters] {10.1093/mnrasl/slw012}, 458, L34

\bibitem[\protect\citeauthoryear{Ellison et~al.,}{Ellison et~al.}{2021}]{Ellison2021}
Ellison S.~L.,  et~al., 2021, \mn@doi [Monthly Notices of the Royal Astronomical Society: Letters] {10.1093/mnrasl/slab047}, 505, L46

\bibitem[\protect\citeauthoryear{Esin, McClintock  \& Narayan}{Esin et~al.}{1997}]{Esin1997}
Esin A.~A.,  McClintock J.~E.,   Narayan R.,  1997, \mn@doi [The Astrophysical Journal] {10.1086/304829}, 489, 865

\bibitem[\protect\citeauthoryear{Faucher-Gigu{\`{e}}re \& Quataert}{Faucher-Gigu{\`{e}}re \& Quataert}{2012}]{Faucher-Giguere2012}
Faucher-Gigu{\`{e}}re C.~A.,  Quataert E.,  2012, \mn@doi [Monthly Notices of the Royal Astronomical Society] {10.1111/j.1365-2966.2012.21512.x}, 425, 605

\bibitem[\protect\citeauthoryear{Fiacconi, Sijacki  \& Pringle}{Fiacconi et~al.}{2018}]{Fiacconi2018}
Fiacconi D.,  Sijacki D.,   Pringle J.~E.,  2018, \mn@doi [Monthly Notices of the Royal Astronomical Society] {10.1093/mnras/sty893}, 477, 3807

\bibitem[\protect\citeauthoryear{Fiore et~al.,}{Fiore et~al.}{2017}]{Fiore2017}
Fiore F.,  et~al., 2017, \mn@doi [Astronomy & Astrophysics] {10.1051/0004-6361/201629478}, 601, A143

\bibitem[\protect\citeauthoryear{Gaburov \& Nitadori}{Gaburov \& Nitadori}{2011}]{Gaburov2011}
Gaburov E.,  Nitadori K.,  2011, \mn@doi [Monthly Notices of the Royal Astronomical Society] {10.1111/j.1365-2966.2011.18313.x}, 414, 129

\bibitem[\protect\citeauthoryear{Gammie}{Gammie}{1999}]{Gammie1999}
Gammie C.~F.,  1999, \mn@doi [The Astrophysical Journal] {10.1086/312207}, 522, L57

\bibitem[\protect\citeauthoryear{Gammie, Shapiro  \& McKinney}{Gammie et~al.}{2004}]{Gammie2004}
Gammie C.~F.,  Shapiro S.~L.,   McKinney J.~C.,  2004, \mn@doi [The Astrophysical Journal] {10.1086/380996}, 602, 312

\bibitem[\protect\citeauthoryear{Gofford, Reeves, Tombesi, Braito, Turner, Miller  \& Cappi}{Gofford et~al.}{2013}]{Gofford2013}
Gofford J.,  Reeves J.~N.,  Tombesi F.,  Braito V.,  Turner T.~J.,  Miller L.,   Cappi M.,  2013, \mn@doi [Monthly Notices of the Royal Astronomical Society] {10.1093/mnras/sts481}, 430, 60

\bibitem[\protect\citeauthoryear{Greene, Ho  \& Ulvestad}{Greene et~al.}{2006}]{Greene2006}
Greene J.~E.,  Ho L.~C.,   Ulvestad J.~S.,  2006, \mn@doi [The Astrophysical Journal] {10.1086/497905}, 636, 56

\bibitem[\protect\citeauthoryear{Harris et~al.,}{Harris et~al.}{2020}]{Harris2020}
Harris C.~R.,  et~al., 2020, \mn@doi [Nature] {10.1038/s41586-020-2649-2}, 585, 357

\bibitem[\protect\citeauthoryear{Henden, Puchwein, Shen  \& Sijacki}{Henden et~al.}{2018}]{Henden2018}
Henden N.~A.,  Puchwein E.,  Shen S.,   Sijacki D.,  2018, \mn@doi [Monthly Notices of the Royal Astronomical Society] {10.1093/mnras/sty1780}, 479, 5385

\bibitem[\protect\citeauthoryear{Hopkins}{Hopkins}{2015}]{Hopkins2015a}
Hopkins P.~F.,  2015, \mn@doi [Monthly Notices of the Royal Astronomical Society] {10.1093/mnras/stv195}, 450, 53

\bibitem[\protect\citeauthoryear{Hopkins \& Quataert}{Hopkins \& Quataert}{2011}]{Hopkins2011}
Hopkins P.~F.,  Quataert E.,  2011, \mn@doi [Monthly Notices of the Royal Astronomical Society] {10.1111/j.1365-2966.2011.18542.x}, 415, 1027

\bibitem[\protect\citeauthoryear{Hopkins, Kere{\v{s}}, O{\~{n}}orbe, Faucher-Gigu{\`{e}}re, Quataert, Murray  \& Bullock}{Hopkins et~al.}{2014}]{Hopkins2014}
Hopkins P.~F.,  Kere{\v{s}} D.,  O{\~{n}}orbe J.,  Faucher-Gigu{\`{e}}re C.-A.,  Quataert E.,  Murray N.,   Bullock J.~S.,  2014, \mn@doi [Monthly Notices of the Royal Astronomical Society] {10.1093/mnras/stu1738}, 445, 581

\bibitem[\protect\citeauthoryear{Hopkins et~al.,}{Hopkins et~al.}{2018}]{Hopkins2018}
Hopkins P.~F.,  et~al., 2018, \mn@doi [Monthly Notices of the Royal Astronomical Society] {10.1093/mnras/sty1690}, 480, 800

\bibitem[\protect\citeauthoryear{Hopkins et~al.,}{Hopkins et~al.}{2023a}]{Hopkins2023b}
Hopkins P.~F.,  et~al., 2023a, arXiv preprint arXiv:2309.13115

\bibitem[\protect\citeauthoryear{Hopkins et~al.,}{Hopkins et~al.}{2023b}]{Hopkins2023}
Hopkins P.~F.,  et~al., 2023b, \mn@doi [Monthly Notices of the Royal Astronomical Society] {10.1093/mnras/stac3489}, 519, 3154

\bibitem[\protect\citeauthoryear{Hunter}{Hunter}{2007}]{Hunter2007}
Hunter J.~D.,  2007, \mn@doi [Computing in Science & Engineering] {10.1109/MCSE.2007.55}, 9, 90

\bibitem[\protect\citeauthoryear{Hu{\v{s}}ko, Lacey, Schaye, Schaller  \& Nobels}{Hu{\v{s}}ko et~al.}{2022}]{Husko2022}
Hu{\v{s}}ko F.,  Lacey C.~G.,  Schaye J.,  Schaller M.,   Nobels F. S.~J.,  2022, \mn@doi [Monthly Notices of the Royal Astronomical Society] {10.1093/mnras/stac2278}, 516, 3750

\bibitem[\protect\citeauthoryear{Jamrozy, Machalski, Mack  \& Klein}{Jamrozy et~al.}{2005}]{Jamrozy2005}
Jamrozy M.,  Machalski J.,  Mack K.-H.,   Klein U.,  2005, \mn@doi [Astronomy & Astrophysics] {10.1051/0004-6361:20042007}, 433, 467

\bibitem[\protect\citeauthoryear{Jetha, Hardcastle, Ponman  \& Sakelliou}{Jetha et~al.}{2008}]{Jetha2008}
Jetha N.~N.,  Hardcastle M.~J.,  Ponman T.~J.,   Sakelliou I.,  2008, \mn@doi [Monthly Notices of the Royal Astronomical Society] {10.1111/j.1365-2966.2008.13959.x}, 391, 1052

\bibitem[\protect\citeauthoryear{Jung et~al.,}{Jung et~al.}{2022}]{Jung2022}
Jung S.~L.,  et~al., 2022, \mn@doi [Monthly Notices of the Royal Astronomical Society] {10.1093/mnras/stac1622}, 515, 22

\bibitem[\protect\citeauthoryear{Kaviraj et~al.,}{Kaviraj et~al.}{2017}]{Kaviraj2017}
Kaviraj S.,  et~al., 2017, \mn@doi [Monthly Notices of the Royal Astronomical Society] {10.1093/mnras/stx126}, 467, 4739

\bibitem[\protect\citeauthoryear{King \& Pounds}{King \& Pounds}{2015}]{King2015}
King A.,  Pounds K.,  2015, \mn@doi [Annual Review of Astronomy and Astrophysics] {10.1146/annurev-astro-082214-122316}, 53, 115

\bibitem[\protect\citeauthoryear{Kormendy \& Ho}{Kormendy \& Ho}{2013}]{Kormendy2013}
Kormendy J.,  Ho L.~C.,  2013, \mn@doi [Annual Review of Astronomy and Astrophysics] {10.1146/annurev-astro-082708-101811}, 51, 511

\bibitem[\protect\citeauthoryear{Koudmani, Somerville, Sijacki, Bourne, Jiang  \& Profit}{Koudmani et~al.}{2023}]{Koudmani2023}
Koudmani S.,  Somerville R.~S.,  Sijacki D.,  Bourne M.~A.,  Jiang Y.-F.,   Profit K.,  2023, arXiv preprint arXiv:2312.08428

\bibitem[\protect\citeauthoryear{Ku{\'{z}}micz, Jamrozy, Kozie{\l}-Wierzbowska  \& We{\.{z}}gowiec}{Ku{\'{z}}micz et~al.}{2017}]{Kuzmicz2017}
Ku{\'{z}}micz A.,  Jamrozy M.,  Kozie{\l}-Wierzbowska D.,   We{\.{z}}gowiec M.,  2017, \mn@doi [Monthly Notices of the Royal Astronomical Society] {10.1093/mnras/stx1830}, 471, 3806

\bibitem[\protect\citeauthoryear{Lacerda, S{\'{a}}nchez, {Cid Fernandes}, L{\'{o}}pez-Cob{\'{a}}, Espinosa-Ponce  \& Galbany}{Lacerda et~al.}{2020}]{Lacerda2020}
Lacerda E. A.~D.,  S{\'{a}}nchez S.~F.,  {Cid Fernandes} R.,  L{\'{o}}pez-Cob{\'{a}} C.,  Espinosa-Ponce C.,   Galbany L.,  2020, \mn@doi [Monthly Notices of the Royal Astronomical Society] {10.1093/mnras/staa008}, 492, 3073

\bibitem[\protect\citeauthoryear{Lammers, Iyer, Ibarra-Medel, Pacifici, S{\'{a}}nchez, Tacchella  \& Woo}{Lammers et~al.}{2023}]{Lammers2023}
Lammers C.,  Iyer K.~G.,  Ibarra-Medel H.,  Pacifici C.,  S{\'{a}}nchez S.~F.,  Tacchella S.,   Woo J.,  2023, \mn@doi [The Astrophysical Journal] {10.3847/1538-4357/acdd57}, 953, 26

\bibitem[\protect\citeauthoryear{Lanson \& Vila}{Lanson \& Vila}{2008a}]{Lanson2008a}
Lanson N.,  Vila J.-P.,  2008a, \mn@doi [SIAM Journal on Numerical Analysis] {10.1137/S0036142903427718}, 46, 1912

\bibitem[\protect\citeauthoryear{Lanson \& Vila}{Lanson \& Vila}{2008b}]{Lanson2008b}
Lanson N.,  Vila J.-P.,  2008b, \mn@doi [SIAM Journal on Numerical Analysis] {10.1137/S003614290444739X}, 46, 1935

\bibitem[\protect\citeauthoryear{Laor \& Netzer}{Laor \& Netzer}{1989}]{Laor1989}
Laor A.,  Netzer H.,  1989, \mn@doi [Monthly Notices of the Royal Astronomical Society] {10.1093/mnras/238.3.897}, 238, 897

\bibitem[\protect\citeauthoryear{Liang, Durier, Babul, Dav{\'{e}}, Oppenheimer, Katz, Fardal  \& Quinn}{Liang et~al.}{2016}]{Liang2016}
Liang L.,  Durier F.,  Babul A.,  Dav{\'{e}} R.,  Oppenheimer B.~D.,  Katz N.,  Fardal M.,   Quinn T.,  2016, \mn@doi [Monthly Notices of the Royal Astronomical Society] {10.1093/mnras/stv2840}, 456, 4266

\bibitem[\protect\citeauthoryear{Lupi, Haardt, Dotti, Fiacconi, Mayer  \& Madau}{Lupi et~al.}{2016}]{Lupi2016}
Lupi A.,  Haardt F.,  Dotti M.,  Fiacconi D.,  Mayer L.,   Madau P.,  2016, \mn@doi [Monthly Notices of the Royal Astronomical Society] {10.1093/mnras/stv2877}, 456, 2993

\bibitem[\protect\citeauthoryear{Maccarone, Gallo  \& Fender}{Maccarone et~al.}{2003}]{Maccarone2003}
Maccarone T.~J.,  Gallo E.,   Fender R.,  2003, \mn@doi [Monthly Notices of the Royal Astronomical Society] {10.1046/j.1365-8711.2003.07161.x}, 345, L19

\bibitem[\protect\citeauthoryear{Machalski, Chy{\.{z}}y, Stawarz  \& Kozie{\l}}{Machalski et~al.}{2007}]{Machalski2007}
Machalski J.,  Chy{\.{z}}y K.~T.,  Stawarz {\L}.,   Kozie{\l} D.,  2007, \mn@doi [Astronomy & Astrophysics] {10.1051/0004-6361:20066121}, 462, 43

\bibitem[\protect\citeauthoryear{Madau \& Dickinson}{Madau \& Dickinson}{2014}]{Madau2014}
Madau P.,  Dickinson M.,  2014, \mn@doi [Annual Review of Astronomy and Astrophysics] {10.1146/annurev-astro-081811-125615}, 52, 415

\bibitem[\protect\citeauthoryear{Madau, Haardt  \& Dotti}{Madau et~al.}{2014}]{Madau2014a}
Madau P.,  Haardt F.,   Dotti M.,  2014, \mn@doi [The Astrophysical Journal] {10.1088/2041-8205/784/2/L38}, 784, L38

\bibitem[\protect\citeauthoryear{McCarthy, Babul, Bower  \& Balogh}{McCarthy et~al.}{2008}]{McCarthy2008}
McCarthy I.~G.,  Babul A.,  Bower R.~G.,   Balogh M.~L.,  2008, \mn@doi [Monthly Notices of the Royal Astronomical Society] {10.1111/j.1365-2966.2008.13141.x}, 386, 1309

\bibitem[\protect\citeauthoryear{McCarthy, Schaye, Bird  \& {Le Brun}}{McCarthy et~al.}{2017}]{McCarthy2017}
McCarthy I.~G.,  Schaye J.,  Bird S.,   {Le Brun} A. M.~C.,  2017, \mn@doi [Monthly Notices of the Royal Astronomical Society] {10.1093/mnras/stw2792}, 465, 2936

\bibitem[\protect\citeauthoryear{McClintock, Shafee, Narayan, Remillard, Davis  \& Li}{McClintock et~al.}{2006}]{McClintock2006}
McClintock J.~E.,  Shafee R.,  Narayan R.,  Remillard R.~A.,  Davis S.~W.,   Li L.,  2006, \mn@doi [The Astrophysical Journal] {10.1086/508457}, 652, 518

\bibitem[\protect\citeauthoryear{Merlin et~al.,}{Merlin et~al.}{2019}]{Merlin2019}
Merlin E.,  et~al., 2019, \mn@doi [Monthly Notices of the Royal Astronomical Society] {10.1093/mnras/stz2615}, 490, 3309

\bibitem[\protect\citeauthoryear{Merloni \& Heinz}{Merloni \& Heinz}{2008}]{Merloni2008}
Merloni A.,  Heinz S.,  2008, \mn@doi [Monthly Notices of the Royal Astronomical Society] {10.1111/j.1365-2966.2008.13472.x}, 388, 1011

\bibitem[\protect\citeauthoryear{Morganti}{Morganti}{2017}]{Morganti2017}
Morganti R.,  2017, \mn@doi [Frontiers in Astronomy and Space Sciences] {10.3389/fspas.2017.00042}, 4, 1

\bibitem[\protect\citeauthoryear{Murray, Quataert  \& Thompson}{Murray et~al.}{2005}]{Murray2005}
Murray N.,  Quataert E.,   Thompson T.~A.,  2005, \mn@doi [The Astrophysical Journal] {10.1086/426067}, 618, 569

\bibitem[\protect\citeauthoryear{Muzzin et~al.,}{Muzzin et~al.}{2013}]{Muzzin2013}
Muzzin A.,  et~al., 2013, \mn@doi [The Astrophysical Journal] {10.1088/0004-637X/777/1/18}, 777, 18

\bibitem[\protect\citeauthoryear{Naab \& Ostriker}{Naab \& Ostriker}{2017}]{Naab2016}
Naab T.,  Ostriker J.~P.,  2017, \mn@doi [Annual Review of Astronomy and Astrophysics] {10.1146/annurev-astro-081913-040019}, 55, 59

\bibitem[\protect\citeauthoryear{Narayan, Igumenshchev  \& Abramowicz}{Narayan et~al.}{2003}]{Narayan2003}
Narayan R.,  Igumenshchev I.~V.,   Abramowicz M.~A.,  2003, \mn@doi [Publications of the Astronomical Society of Japan] {10.1093/pasj/55.6.L69}, 55, L69

\bibitem[\protect\citeauthoryear{Narayanan et~al.,}{Narayanan et~al.}{2021}]{Narayanan2021}
Narayanan D.,  et~al., 2021, \mn@doi [The Astrophysical Journal Supplement Series] {10.3847/1538-4365/abc487}, 252, 12

\bibitem[\protect\citeauthoryear{Nemmen, Bower, Babul  \& Storchi-Bergmann}{Nemmen et~al.}{2007}]{Nemmen2007}
Nemmen R.~S.,  Bower R.~G.,  Babul A.,   Storchi-Bergmann T.,  2007, \mn@doi [Monthly Notices of the Royal Astronomical Society] {10.1111/j.1365-2966.2007.11726.x}, 377, 1652

\bibitem[\protect\citeauthoryear{Nusser, Silk  \& Babul}{Nusser et~al.}{2006}]{Nusser2006}
Nusser A.,  Silk J.,   Babul A.,  2006, \mn@doi [Monthly Notices of the Royal Astronomical Society] {10.1111/j.1365-2966.2006.11061.x}, 373, 739

\bibitem[\protect\citeauthoryear{O'Sullivan et~al.,}{O'Sullivan et~al.}{2012}]{OSullivan2012}
O'Sullivan E.,  et~al., 2012, \mn@doi [Monthly Notices of the Royal Astronomical Society] {10.1111/j.1365-2966.2012.21459.x}, 424, 2971

\bibitem[\protect\citeauthoryear{Oppenheimer, Babul, Bah{\'{e}}, Butsky  \& McCarthy}{Oppenheimer et~al.}{2021}]{Oppenheimer2021}
Oppenheimer B.~D.,  Babul A.,  Bah{\'{e}} Y.,  Butsky I.~S.,   McCarthy I.~G.,  2021, \mn@doi [Universe] {10.3390/universe7070209}, 7, 209

\bibitem[\protect\citeauthoryear{Perez \& Granger}{Perez \& Granger}{2007}]{Perez2007}
Perez F.,  Granger B.~E.,  2007, \mn@doi [Computing in Science & Engineering] {10.1109/MCSE.2007.53}, 9, 21

\bibitem[\protect\citeauthoryear{Pillepich et~al.,}{Pillepich et~al.}{2018}]{Pillepich2018}
Pillepich A.,  et~al., 2018, \mn@doi [Monthly Notices of the Royal Astronomical Society] {10.1093/mnras/stx3112}, 475, 648

\bibitem[\protect\citeauthoryear{{Planck Collaboration XIII}}{{Planck Collaboration XIII}}{2015}]{Ade2016}
{Planck Collaboration XIII} 2015, \mn@doi [Astronomy & Astrophysics] {10.1051/0004-6361/201525830}, 594, A13

\bibitem[\protect\citeauthoryear{Prasad, Sharma  \& Babul}{Prasad et~al.}{2015}]{Prasad2015}
Prasad D.,  Sharma P.,   Babul A.,  2015, \mn@doi [The Astrophysical Journal] {10.1088/0004-637X/811/2/108}, 811, 108

\bibitem[\protect\citeauthoryear{Prasad, Sharma  \& Babul}{Prasad et~al.}{2018}]{Prasad2018}
Prasad D.,  Sharma P.,   Babul A.,  2018, \mn@doi [The Astrophysical Journal] {10.3847/1538-4357/aacce8}, 863, 62

\bibitem[\protect\citeauthoryear{Qiu, Bogdanovi{\'{c}}, Li, Park  \& Wise}{Qiu et~al.}{2019}]{Qiu2019}
Qiu Y.,  Bogdanovi{\'{c}} T.,  Li Y.,  Park K.,   Wise J.~H.,  2019, \mn@doi [The Astrophysical Journal] {10.3847/1538-4357/ab18fd}, 877, 47

\bibitem[\protect\citeauthoryear{Reynolds, McKernan, Fabian, Stone  \& Vernaleo}{Reynolds et~al.}{2005}]{Reynolds2005}
Reynolds C.~S.,  McKernan B.,  Fabian A.~C.,  Stone J.~M.,   Vernaleo J.~C.,  2005, \mn@doi [Monthly Notices of the Royal Astronomical Society] {10.1111/j.1365-2966.2005.08643.x}, 357, 242

\bibitem[\protect\citeauthoryear{Reynolds, Brenneman, Lohfink, Trippe, Miller, Reis, Nowak  \& Fabian}{Reynolds et~al.}{2012}]{Reynolds2012}
Reynolds C.~S.,  Brenneman L.~W.,  Lohfink A.~M.,  Trippe M.~L.,  Miller J.~M.,  Reis R.~C.,  Nowak M.~A.,   Fabian A.~C.,  2012, in AIP Conference Proceedings. pp 157--164, \mn@doi{10.1063/1.3696170}, \url {https://pubs.aip.org/aip/acp/article/1427/1/157-164/837608}

\bibitem[\protect\citeauthoryear{Ricarte, Narayan  \& Curd}{Ricarte et~al.}{2023}]{Ricarte2023}
Ricarte A.,  Narayan R.,   Curd B.,  2023, \mn@doi [The Astrophysical Journal Letters] {10.3847/2041-8213/aceda5}, 954, L22

\bibitem[\protect\citeauthoryear{Rosas-Guevara et~al.,}{Rosas-Guevara et~al.}{2015}]{Rosas-Guevara2015a}
Rosas-Guevara Y.~M.,  et~al., 2015, \mn@doi [Monthly Notices of the Royal Astronomical Society] {10.1093/mnras/stv2056}, 454, 1038

\bibitem[\protect\citeauthoryear{Ruszkowski \& Oh}{Ruszkowski \& Oh}{2011}]{Ruszkowski2011}
Ruszkowski M.,  Oh S.~P.,  2011, \mn@doi [Monthly Notices of the Royal Astronomical Society] {10.1111/j.1365-2966.2011.18482.x}, 414, 1493

\bibitem[\protect\citeauthoryear{S{\c{a}}dowski}{S{\c{a}}dowski}{2009}]{Sadowski2009}
S{\c{a}}dowski A.,  2009, \mn@doi [The Astrophysical Journal Supplement Series] {10.1088/0067-0049/183/2/171}, 183, 171

\bibitem[\protect\citeauthoryear{S{\c{a}}dowski, Narayan, McKinney  \& Tchekhovskoy}{S{\c{a}}dowski et~al.}{2014}]{Sadowski2014}
S{\c{a}}dowski A.,  Narayan R.,  McKinney J.~C.,   Tchekhovskoy A.,  2014, \mn@doi [Monthly Notices of the Royal Astronomical Society] {10.1093/mnras/stt2479}, 439, 503

\bibitem[\protect\citeauthoryear{Sala, Cenci, Capelo, Lupi  \& Dotti}{Sala et~al.}{2020}]{Sala2020}
Sala L.,  Cenci E.,  Capelo P.~R.,  Lupi A.,   Dotti M.,  2020, \mn@doi [Monthly Notices of the Royal Astronomical Society] {10.1093/mnras/staa3552}, 500, 4788

\bibitem[\protect\citeauthoryear{Salim et~al.,}{Salim et~al.}{2016}]{Salim2016}
Salim S.,  et~al., 2016, \mn@doi [The Astrophysical Journal Supplement Series] {10.3847/0067-0049/227/1/2}, 227, 2

\bibitem[\protect\citeauthoryear{Salim, Boquien  \& Lee}{Salim et~al.}{2018}]{Salim2018}
Salim S.,  Boquien M.,   Lee J.~C.,  2018, \mn@doi [The Astrophysical Journal] {10.3847/1538-4357/aabf3c}, 859, 11

\bibitem[\protect\citeauthoryear{Sanchez et~al.,}{Sanchez et~al.}{2018}]{Sanchez2018}
Sanchez S.~F.,  et~al., 2018, Revista Mexicana de Astronomia y Astrofisica, 54, 217

\bibitem[\protect\citeauthoryear{Sazonov, Ostriker, Ciotti  \& Sunyaev}{Sazonov et~al.}{2005}]{Sazonov2005}
Sazonov S.~Y.,  Ostriker J.~P.,  Ciotti L.,   Sunyaev R.~A.,  2005, \mn@doi [Monthly Notices of the Royal Astronomical Society] {10.1111/J.1365-2966.2005.08763.x}, 358, 168

\bibitem[\protect\citeauthoryear{Schaye et~al.,}{Schaye et~al.}{2015}]{Schaye2014}
Schaye J.,  et~al., 2015, \mn@doi [Monthly Notices of the Royal Astronomical Society] {10.1093/mnras/stu2058}, 446, 521

\bibitem[\protect\citeauthoryear{Schaye et~al.,}{Schaye et~al.}{2023}]{Schaye2023}
Schaye J.,  et~al., 2023, \mn@doi [Monthly Notices of the Royal Astronomical Society] {10.1093/mnras/stad2419}, 526, 4978

\bibitem[\protect\citeauthoryear{Scheuer \& Feiler}{Scheuer \& Feiler}{1996}]{Scheuer1996}
Scheuer P.,  Feiler R.,  1996, \mn@doi [Monthly Notices of the Royal Astronomical Society] {10.1093/mnras/282.1.291}, 282, 291

\bibitem[\protect\citeauthoryear{Schutte, Reines  \& Greene}{Schutte et~al.}{2019}]{Schutte2019}
Schutte Z.,  Reines A.~E.,   Greene J.~E.,  2019, \mn@doi [The Astrophysical Journal] {10.3847/1538-4357/ab35dd}, 887, 245

\bibitem[\protect\citeauthoryear{Shakura \& Sunyaev}{Shakura \& Sunyaev}{1973}]{Shakura1973}
Shakura N.~I.,  Sunyaev R.~a.,  1973, \mn@doi [Astronomy and Astrophysics] {10.1086/170270}, 24, 337

\bibitem[\protect\citeauthoryear{Sijacki, Springel, {Di Matteo}  \& Hernquist}{Sijacki et~al.}{2007}]{Sijacki2007}
Sijacki D.,  Springel V.,  {Di Matteo} T.,   Hernquist L.,  2007, \mn@doi [Monthly Notices of the Royal Astronomical Society] {10.1111/j.1365-2966.2007.12153.x}, 380, 877

\bibitem[\protect\citeauthoryear{Sijacki, Vogelsberger, Genel, Springel, Torrey, Snyder, Nelson  \& Hernquist}{Sijacki et~al.}{2015}]{Sijacki2015a}
Sijacki D.,  Vogelsberger M.,  Genel S.,  Springel V.,  Torrey P.,  Snyder G.~F.,  Nelson D.,   Hernquist L.,  2015, \mn@doi [Monthly Notices of the Royal Astronomical Society] {10.1093/mnras/stv1340}, 452, 575

\bibitem[\protect\citeauthoryear{Silk \& Rees}{Silk \& Rees}{1998}]{Silk1998}
Silk J.,  Rees M.~J.,  1998, Monthly Notices of the Royal Astronomical Society, 324, 128

\bibitem[\protect\citeauthoryear{Smith, Mushotzky, Vogel, Shimizu  \& Miller}{Smith et~al.}{2016}]{Smith2016}
Smith K.~L.,  Mushotzky R.~F.,  Vogel S.,  Shimizu T.~T.,   Miller N.,  2016, \mn@doi [The Astrophysical Journal] {10.3847/0004-637X/832/2/163}, 832, 163

\bibitem[\protect\citeauthoryear{So{\l}tan}{So{\l}tan}{1982}]{Soltan1982}
So{\l}tan A.,  1982, \mn@doi [Monthly Notices of the Royal Astronomical Society] {10.1093/mnras/200.1.115}, 200, 115

\bibitem[\protect\citeauthoryear{Somerville \& Dav{\'{e}}}{Somerville \& Dav{\'{e}}}{2015}]{Somerville2015}
Somerville R.~S.,  Dav{\'{e}} R.,  2015, \mn@doi [Annu. Rev. Astron. Astrophys] {10.1146/annurev-astro-082812-140951}, 53, 51

\bibitem[\protect\citeauthoryear{Song et~al.,}{Song et~al.}{2016}]{Song2016}
Song M.,  et~al., 2016, \mn@doi [The Astrophysical Journal] {10.3847/0004-637X/825/1/5}, 825, 5

\bibitem[\protect\citeauthoryear{Springel, {Di Matteo}  \& Hernquist}{Springel et~al.}{2005}]{Springel2005b}
Springel V.,  {Di Matteo} T.,   Hernquist L.,  2005, \mn@doi [Monthly Notices of the Royal Astronomical Society] {10.1111/j.1365-2966.2005.09238.x}, 361, 776

\bibitem[\protect\citeauthoryear{Su et~al.,}{Su et~al.}{2021}]{Su2021}
Su K.-Y.,  et~al., 2021, \mn@doi [Monthly Notices of the Royal Astronomical Society] {10.1093/mnras/stab2021}, 507, 175

\bibitem[\protect\citeauthoryear{Talbot, Bourne  \& Sijacki}{Talbot et~al.}{2021}]{Talbot2020}
Talbot R.~Y.,  Bourne M.~A.,   Sijacki D.,  2021, \mn@doi [Monthly Notices of the Royal Astronomical Society] {10.1093/mnras/stab804}, 504, 3619

\bibitem[\protect\citeauthoryear{Talbot, Sijacki  \& Bourne}{Talbot et~al.}{2022}]{Talbot2022}
Talbot R.~Y.,  Sijacki D.,   Bourne M.~A.,  2022, \mn@doi [Monthly Notices of the Royal Astronomical Society] {10.1093/mnras/stac1566}, 514, 4535

\bibitem[\protect\citeauthoryear{Talbot, Sijacki  \& Bourne}{Talbot et~al.}{2024}]{Talbot2024}
Talbot R.~Y.,  Sijacki D.,   Bourne M.~A.,  2024, \mn@doi [Monthly Notices of the Royal Astronomical Society] {10.1093/mnras/stae392}, 528, 5432

\bibitem[\protect\citeauthoryear{Tchekhovskoy, McKinney  \& Narayan}{Tchekhovskoy et~al.}{2012}]{Tchekhovskoy2012}
Tchekhovskoy A.,  McKinney J.~C.,   Narayan R.,  2012, \mn@doi [Journal of Physics: Conference Series] {10.1088/1742-6596/372/1/012040}, 372, 012040

\bibitem[\protect\citeauthoryear{Thorne}{Thorne}{1974}]{Thorne1974}
Thorne K.~S.,  1974, \mn@doi [The Astrophysical Journal] {10.1086/152991}, 191, 507

\bibitem[\protect\citeauthoryear{Tombesi, Cappi, Reeves, Palumbo, Yaqoob, Braito  \& Dadina}{Tombesi et~al.}{2010}]{Tombesi2010}
Tombesi F.,  Cappi M.,  Reeves J.~N.,  Palumbo G. G.~C.,  Yaqoob T.,  Braito V.,   Dadina M.,  2010, \mn@doi [Astronomy and Astrophysics] {10.1051/0004-6361/200913440}, 521, A57

\bibitem[\protect\citeauthoryear{Tombesi, Cappi, Reeves, Palumbo, Braito  \& Dadina}{Tombesi et~al.}{2011}]{Tombesi2011}
Tombesi F.,  Cappi M.,  Reeves J.~N.,  Palumbo G. G.~C.,  Braito V.,   Dadina M.,  2011, \mn@doi [The Astrophysical Journal] {10.1088/0004-637X/742/1/44}, 742, 44

\bibitem[\protect\citeauthoryear{Tombesi, Tazaki, Mushotzky, Ueda, Cappi, Gofford, Reeves  \& Guainazzi}{Tombesi et~al.}{2014}]{Tombesi2014}
Tombesi F.,  Tazaki F.,  Mushotzky R.~F.,  Ueda Y.,  Cappi M.,  Gofford J.,  Reeves J.~N.,   Guainazzi M.,  2014, \mn@doi [Monthly Notices of the Royal Astronomical Society] {10.1093/mnras/stu1297}, 443, 2154

\bibitem[\protect\citeauthoryear{Tomczak et~al.,}{Tomczak et~al.}{2014}]{Tomczak2014}
Tomczak A.~R.,  et~al., 2014, \mn@doi [The Astrophysical Journal] {10.1088/0004-637X/783/2/85}, 783, 85

\bibitem[\protect\citeauthoryear{Tremmel, Karcher, Governato, Volonteri, Quinn, Pontzen, Anderson  \& Bellovary}{Tremmel et~al.}{2017}]{Tremmel2017a}
Tremmel M.,  Karcher M.,  Governato F.,  Volonteri M.,  Quinn T.~R.,  Pontzen A.,  Anderson L.,   Bellovary J.,  2017, \mn@doi [Monthly Notices of the Royal Astronomical Society] {10.1093/mnras/stx1160}, 470, 1121

\bibitem[\protect\citeauthoryear{Turk, Smith, Oishi, Skory, Skillman, Abel  \& Norman}{Turk et~al.}{2011}]{Turk2011}
Turk M.~J.,  Smith B.~D.,  Oishi J.~S.,  Skory S.,  Skillman S.~W.,  Abel T.,   Norman M.~L.,  2011, \mn@doi [The Astrophysical Journal Supplement Series] {10.1088/0067-0049/192/1/9}, 192, 9

\bibitem[\protect\citeauthoryear{Virtanen et~al.,}{Virtanen et~al.}{2020}]{Virtanen2020}
Virtanen P.,  et~al., 2020, \mn@doi [Nature Methods] {10.1038/s41592-019-0686-2}, 17, 261

\bibitem[\protect\citeauthoryear{Vogelsberger et~al.,}{Vogelsberger et~al.}{2014}]{Vogelsberger2014}
Vogelsberger M.,  et~al., 2014, \mn@doi [Monthly Notices of the Royal Astronomical Society] {10.1093/mnras/stu1536}, 444, 1518

\bibitem[\protect\citeauthoryear{Weinberger et~al.,}{Weinberger et~al.}{2017}]{Weinberger2016}
Weinberger R.,  et~al., 2017, \mn@doi [Monthly Notices of the Royal Astronomical Society] {10.1093/mnras/stw2944}, 465, 3291

\bibitem[\protect\citeauthoryear{Weinberger et~al.,}{Weinberger et~al.}{2023}]{Weinberger2023}
Weinberger R.,  et~al., 2023, \mn@doi [Monthly Notices of the Royal Astronomical Society] {10.1093/mnras/stad1396}, 523, 1104

\bibitem[\protect\citeauthoryear{Wellons et~al.,}{Wellons et~al.}{2023}]{Wellons2023}
Wellons S.,  et~al., 2023, \mn@doi [Monthly Notices of the Royal Astronomical Society] {10.1093/mnras/stad511}, 520, 5394

\bibitem[\protect\citeauthoryear{Yuan \& Narayan}{Yuan \& Narayan}{2014}]{Yuan2014}
Yuan F.,  Narayan R.,  2014, \mn@doi [Annual Review of Astronomy and Astrophysics] {10.1146/annurev-astro-082812-141003}, 52, 529

\bibitem[\protect\citeauthoryear{Zubovas \& King}{Zubovas \& King}{2012}]{Zubovas2012}
Zubovas K.,  King A.,  2012, \mn@doi [The Astrophysical Journal] {10.1088/2041-8205/745/2/L34}, 745, L34

\makeatother
\end{thebibliography}




\bsp	
\label{lastpage}
\end{document}